\documentclass[journal,10pt]{IEEEtran}
\usepackage{amsmath}
\usepackage{graphicx}
\usepackage{indentfirst}
\usepackage{cases}
\usepackage{cite}
\usepackage{filecontents}
\usepackage{tabularx,booktabs, caption}
\usepackage{makecell}

\newcommand{\blue}{\textcolor{black}}

\newcolumntype{C}{>{\centering\arraybackslash}X} % centered version of "X" type
\usepackage{lipsum}
\usepackage{fancyhdr}
\usepackage{amsmath,amssymb,mathtools,bm,etoolbox}
\usepackage{color}
\usepackage{array}
\usepackage{mathtools}
\usepackage[british]{babel}
\usepackage{csquotes}
\usepackage{nccmath}
\usepackage{gensymb}
\usepackage[shortlabels]{enumitem}
\usepackage[nodisplayskipstretch]{setspace}
\usepackage[ruled,linesnumbered]{algorithm2e}
\usepackage{comment}
\usepackage{amsthm}
\newtheorem{theorem}{Theorem}

\newtheorem{proposition}[theorem]{Proposition}

\newtheorem{remark}{Remark}
\let\emptyset\varnothing
\SetLabelAlign{bibright}{\hss\llap{[#1]}}
\newcounter{mynum}

\usepackage{bbold}
\usepackage{hyperref}
\hypersetup{
     colorlinks   = true,
     citecolor    = red,
     linkcolor    = red,
     urlcolor     = black
}

\allowdisplaybreaks % Break equation
\newcommand{\etal}{\emph{et al. }}
\usepackage{multirow}
\usepackage{balance}

\makeatletter 
\pretocmd\@bibitem{\color{black}\csname keycolor#1\endcsname}{}{\fail}
\newcommand\citecolor[1]{\@namedef{keycolor#1}{\color{black}}}
\makeatother
\usepackage{xcolor,cite,etoolbox}
\citecolor{11481114}
\citecolor{albataineh2025adaptive}
\citecolor{11422871}
\citecolor{3gpp36777}
\citecolor{6863654}
\title{\huge Mobility- and Feedback-Aware Multi-Level Conflict-Triggered Hybrid Beamforming for Multi-User mmWave UAV Systems}
\begin{document}
\author{Thuan Van Le, 
Nguyen Cong Luong,
Xingwang Li,~\IEEEmembership{Senior Member, IEEE}, \\
Vo Nguyen Quoc Bao,~\IEEEmembership{Senior Member,~IEEE},
and Ngo~Hoang~Tu,~\IEEEmembership{Member,~IEEE}
\vspace{-0.25cm}
    \thanks{T. V. Le is with the Faculty of Electrical and Electronic Engineering, Phenikaa School of Engineering, Phenikaa University, Hanoi 12116, Vietnam (e-mail: thuan.levan@phenikaa-uni.edu.vn).}
    \thanks{N. C. Luong is with the Phenikaa School of Computing, Phenikaa University, Hanoi 12116, Vietnam (e-mail: luong.nguyencong@phenikaa-uni.edu.vn).}
    \thanks{X. Li is with the School of Physics and Electronic Information Engineering, Henan Polytechnic University, Jiaozuo 454003, China (e-mail: lixingwangbupt@gmail.com).}
    \thanks{Vo Nguyen Quoc Bao and Ngo Hoang Tu are with the Faculty of Information Technology, Van Lang School of Technology, Van Lang University, Ho Chi Minh City 70000, Vietnam (e-mail: bao.vnq@vlu.edu.vn, tu.nh@vlu.edu.vn). \textit{(Corresponding author: Ngo Hoang Tu.)}}
    }%
\maketitle
%\thispagestyle{firstpage}

% \markboth{Manuscript Submitted to IEEE Transactions on Communications, 2026}{}

\begin{abstract}
This paper investigates hybrid beamforming for multi-user large multiple-input multiple-output millimeter-wave unmanned aerial vehicle (UAV) downlink systems under mobility-induced channel aging and delayed beam-training feedback.
\blue{Analog beam selection from compact delayed reports is a partial-observation decision, while additional candidate evaluations consume processing time and reduce the useful payload interval.}
We propose a mobility- and feedback-aware multi-level refinement strategy, termed MLR-TG, to improve robustness without always-on candidate search.
\blue{Candidate subsets are ranked by a predicted net utility constructed from quantized complex coefficients of the reported codewords and the UAV mobility state, while the transmission regularized zero-forcing precoder is computed once from pilot-estimated effective channel state information (CSI) after analog selection.}
The refinement level is adaptively selected according to conflict severity and aging sensitivity.
\blue{The selection rule is a two-statistic approximation of predicted-utility maximization, employs a system-size-invariant conflict score, and is calibrated offline on training data disjoint from evaluation.}
\blue{Simulations on a three-dimensional air-to-ground model with UAV attitude dynamics and common channel trajectories show that MLR-TG reduces system outage probability by 26.7\% and improves the 5th-percentile user rate by 53.9\% relative to greedy sector beamforming, while net spectral efficiency remains within 0.96\%. Compared with always-on global top-$3$ refinement, MLR-TG improves net spectral efficiency by 5.5\% while evaluating 77.9\% fewer candidates, and remains within 3.4\% of a noncausal-CSI level oracle in net spectral efficiency while requiring 86.9\% fewer feedback bits than full-CSI reporting.}
\end{abstract}


\vspace{-0.2cm}
\section{Introduction}
\label{sec:introduction}

% \textcolor{red}{Section~\ref{sec:introduction} and Abstract will be checked at the final stage...}

\subsection{Research Context and Literature Overview}
\label{Sec:IA}

Millimeter-wave (mmWave) multiple-input multiple-output (MIMO) communications have become a key enabler for high spectral efficiency (SE) wireless systems in fifth-generation and beyond, owing to the availability of wide bandwidth and highly directional transmission \cite{11224256}.
% \cite{10623156,11224256,8373698}. 
However, fully digital beamforming in mmWave MIMO systems is often impractical because each antenna requires a dedicated radio-frequency (RF) chain, resulting in high hardware cost and power consumption \cite{9513786,9149126}. 
Hybrid beamforming (HBF) has therefore emerged as a practical solution by combining high-dimensional analog pre/post-processing with lower-dimensional digital processing, achieving a favorable balance among beamforming performance, energy efficiency (EE), and implementation complexity \cite{6717211,7389996}.
% 8932399

This trade-off becomes more pronounced in unmanned aerial vehicle (UAV)-assisted mmWave systems \cite{dao2021survey,8709739}. 
Compared with terrestrial links, UAV communications benefit from stronger line-of-sight (LoS) propagation, flexible deployment, and improved spatial coverage, making mmWave transmission attractive for aerial high-rate links \cite{8288376,wang2020beamforming}. 
However, UAV mobility introduces additional challenges, including Doppler-induced channel aging, beam misalignment, beam conflict, and rapidly varying angular geometry.
These factors can significantly degrade beamforming performance if beam management is not conservatively designed \cite{8334304,10770152,8811740}. 
Therefore, HBF in mmWave multi-user (MU) MIMO UAV systems must jointly account for beamforming gain, limited feedback, mobility sensitivity, beam conflict, and computational complexity \cite{10278791}.
Existing works related to this problem can be broadly categorized into three directions.

%------------------------------
\textit{1) Full Channel State Information (CSI)-Based HBF:}
Prior work on mmWave MIMO UAV systems often assumes full or high-quality CSI, enabling optimization-based designs for advanced HBF.
For instance, Chen \etal \cite{9658041} studied joint UAV altitude control and HBF in mmWave MU massive MIMO systems, achieving near fully digital performance via weighted minimum mean squared error-based alternating optimization (AO), while cooperative AO with residual error minimization \cite{wang2023cooperative} and coordinated descent over coupled precoder problems \cite{mir2022relay} were developed for single-user (SU) MIMO and wideband SU large MIMO UAV links, respectively.
In addition, Gao \etal \cite{gao2022data} proposed a data-driven HBF framework for MU massive multiple-input single-output (MISO) systems, leveraging implicit CSI through end-to-end learning to reduce pilot and feedback overhead while maintaining high SE.
Despite their strong performance, these approaches typically rely on accurate or implicitly learned CSI and involve computationally intensive optimization or training procedures.
As such, they do not explicitly address beam conflict resolution and are less suitable for UAV scenarios with rapidly time-varying channels, limited feedback, and stringent computational constraints \cite{4641946}.

% pp1
% For example, Zhang \etal \cite{zhang2022adaptive} developed an adaptive codebook-based channel estimation scheme with HBF that exploits full-CSI multi-carrier observations to improve estimation accuracy, particularly in low signal-to-noise ratio (SNR) regimes.
% % pp2
% In \cite{ge2022training}, Ge \etal proposed a training beam strategy via compressive sensing and HBF optimization to enhance channel recovery performance.
% %
% Meanwhile, the authors in \cite{tu2026hybrid} and \cite{tu2025semi}...
% In \cite{tu2026multi}...
% sparse precoding, hybrid beamformer decomposition, and channel estimation for large antenna arrays \cite{6717211,7389996,6847111}. 
% especially in mmWave massive MIMO settings \cite{10547480,9658041}. Such overhead becomes even less attractive in UAV scenarios, where the channel varies quickly and frequent CSI acquisition is difficult to maintain \cite{4641946,7445130,8334304}.

%-------------------------------------
\textit{2) UAV Mobility-Aware mmWave Beam Management:}
% The second direction concerns UAV-specific mmWave communications, where prior works have emphasized LoS-dominant sparse channels, directional transmission, and beam management under aerial mobility \cite{8288376,wang2020beamforming,8709739,9390407}. These studies show that Doppler-induced channel aging, beam misalignment, and rapidly varying angular geometry can severely affect link reliability if the beam management procedure does not account for mobility \cite{8334304,10770152,8811740,9613775,10356727}. Therefore, for multi-user mmWave UAV systems, beamforming design must jointly consider not only beamforming gain but also mobility sensitivity, feedback delay, and online complexity \cite{10278791}.
Prior work in this direction has focused on mobility-aware beam management and channel dynamics in UAV communications.
% pp8
Specifically, Wang \etal \cite{wang2020beamforming} analyzed the trade-off between analog and HBF in mmWave SU massive MISO UAV networks, highlighting the impact of UAV mobility on coverage area, EE, and SE.
% pp5
In UAV-assisted MU mmWave MISO systems, Zhao \etal \cite{8334304} proposed a channel tracking framework that integrates flight control information with channel estimation to capture mobility-induced variations.
% pp9
In \cite{8811740}, beam tracking under UAV mobility was further investigated in MU massive MISO systems, where adaptive beam training and angular velocity estimation were developed to adjust training frequency and beamwidth according to channel dynamics.
% pp10
\blue{In addition, Hyun \etal \cite{hyun2021adaptive} designed a mobility-aware beam strategy for vehicle-to-infrastructure links, where extended Kalman filtering-based tracking was employed to enable reliable beam alignment with reduced feedback overhead in high-mobility scenarios, and the same filtering principle transfers to aerial platforms once the state model is replaced}.
\blue{Robustness of the training stage itself has also been addressed, through soft-decision beam training that accounts for measurement reliability \cite{11481114}, through adaptive beam pairing with local interpolation for wideband dictionaries \cite{albataineh2025adaptive}, and through parametric a priori-aided beamspace tracking that predicts the dominant taps instead of re-acquiring them \cite{11422871}.}
%Survey
While these works effectively address mobility-induced channel variation and beam alignment, they primarily focus on single-link reliability and tracking accuracy, without explicitly considering MU beam conflict or adaptive candidate refinement under limited feedback.
% survey \cite{li2022mobility}.

% \textcolor{red}{Check until this...}

%-----------------------
\textit{3) Codebook-Based Beam Selection and Limited-Feedback Beam Management:}
This line of work considers codebook-based beam training and limited-feedback beam management as practical alternatives to full CSI acquisition \cite{8458146}.
In-field measurements for UAV beamforming in SU MISO systems quantified the impact of feedback overhead, channel reciprocity, and update rate under mobility \cite{shi2018uabeam}, and hybrid feedback schemes were subsequently proposed to improve throughput under mobility-induced variations \cite{shi2021channel}.
Efficient quantization with full diversity \cite{rajashekar2018beamforming} and Kalman filtering against synchronization and estimation errors \cite{hanna2022distributed} were also developed for low-altitude air-to-ground links.
In these approaches, the transmitter relies on compact beam indicators, ranked beam lists, or codebook-based training results instead of reconstructing the full instantaneous channel, which reduces feedback overhead and simplifies online beam selection.

When codebook-based beam selection is extended to MU settings, beam selection becomes more challenging due to user equipment (UE) contention over limited angular resources.
Greedy and low-complexity beam selection methods are appealing due to their implementation simplicity.
Specifically, Singh \etal \cite{10199880} developed greedy algorithms for joint UE and beam selection in MU MISO mmWave systems to improve weighted sum-SE under MU contention, and Murshed \etal \cite{10480178} proposed an interference-aware greedy beam selection algorithm for MU massive MIMO mmWave systems, achieving a favorable trade-off between SE and computational complexity.
However, such greedy approaches typically rely on instantaneous selection rules and lack robustness when UEs compete for similar angular directions, request overlapping beams, or exhibit ambiguous beam preferences. In such cases, purely greedy beam assignment may suffer from insufficient beam diversity and degraded interference management.

\vspace{-0.25cm}
%----------
\subsection{Novelty and Contributions}
The discussions in Section~\ref{Sec:IA} reveal a fundamental gap that is not adequately addressed by the existing literature. 
Full-CSI HBF methods incur substantial signaling and computational overhead under UAV mobility.
Mobility-aware beam tracking approaches capture channel aging and beam misalignment, but do not address adaptive MU beam selection from compact feedback.
Meanwhile, codebook-based and greedy beam selection methods operate under limited feedback but rely on fixed selection rules, without adapting the refinement depth to the joint effects of beam conflict, mobility-induced aging, and feedback delay.
%
% In particular, once candidate refinement is invoked, existing methods typically adopt a fixed refinement depth, regardless of the operating conditions. However,
It is worth mentioning that the optimal refinement depth should depend not only on the severity of beam conflict, but also on the expected training-transmission mismatch induced by UAV motion and beam-management latency. Ignoring this coupling leads to either insufficient beam exploration under severe conflict or excessive processing delay under strong aging.

% ; prior UAV beam tracking studies highlight channel aging but do not address adaptive multi-user candidate refinement from compact beam feedback; and prior codebook-based or greedy beam selection methods do not explicitly adapt the refinement depth to the joint effect of beam conflict, mobility-induced aging, and feedback delay. In particular, once candidate refinement is introduced, the most suitable refinement depth should depend not only on the severity of beam conflict, but also on the expected training--transmission mismatch caused by UAV motion and beam-management latency. This motivates the development of a mobility- and feedback-aware multi-level conflict-triggered HBF framework for multi-user mmWave UAV systems.

To address this limitation, we propose a mobility- and feedback-aware multi-level conflict-triggered hybrid beamforming framework for mmWave UAV communications.
The key idea is to extend the conventional single-level triggered refinement into a multi-level adaptive policy. 
Specifically, we first construct a foundational single-level design in which the greedy sector (GS) solution serves as the default low-complexity mode.
We then establish triggered global top-3 refinement (TG3-TR) as a baseline and generalize it into a multi-level refinement policy, termed multi-level refinement for GS beamforming (MLR-TG).
Instead of making a binary decision on whether to activate refinement, the proposed framework dynamically selects the refinement depth based on both the beam conflict structure revealed by compact beam-training feedback and the expected severity of mobility-induced aging.
As a result, the transmitter can switch among GS beamforming, sector top-2 refinement, global top-2 refinement, and global top-3 refinement according to the current operating conditions. This enables the system to avoid unnecessarily deep candidate search under severe aging, while still exploiting richer candidate exploration when beam conflict is significant and the expected mismatch remains moderate.
% The key idea is to establish triggered global top-$3$ refinement (TG3-TR) as a single-level precursor and then generalize it to a multi-level refinement policy, termed multi-level refinement for GS beamforming (MLR-TG), that adaptively selects the refinement depth according to both the conflict structure revealed by compact beam-training feedback and the expected severity of mobility-induced aging. Instead of making only a binary decision on whether refinement should be activated, the proposed framework allows the transmitter to choose among the GS baseline, a sector-top-$2$ refinement, a global top-$2$ refinement, and a global top-$3$ refinement. In this way, the transmitter can avoid unnecessarily deep candidate search under strong aging, while still exploiting richer candidate exploration when the beam conflict becomes severe and the expected mismatch remains moderate.

% \textcolor{red}{Insert a comparison table here...}

\begin{table*}[t]\color{black}
% \vspace{-0.5cm}
\centering
\caption{\color{black}Comparison between the proposed and related works.}
\label{tab:novelty_comparison}
\renewcommand{\arraystretch}{1.1}
\resizebox{\linewidth}{!}{
\begin{tabular}{|>{\centering\arraybackslash}p{1.5cm}|>{\centering\arraybackslash}p{2.5cm}|c|c|c|c|c|c|p{11cm}|}
\hline
\multirow{2}{*}{\textbf{Work}}  & \centering \multirow{2}{*}{\textbf{System}} &
\multirow{2}{*}{\textbf{UAV}} & \multirow{2}{*}{\textbf{HBF}} & \textbf{mm-} & \textbf{Limited} &
\multirow{2}{*}{\textbf{Mobility}} & \textbf{Beam} &
\hspace{2.5cm}\multirow{2}{*}{\textbf{Techniques / Contributions}} \\
\textbf{}  & \centering \textbf{} &
\textbf{} & \textbf{} & \textbf{Wave} & \textbf{Feedback} &
\textbf{} & \textbf{Conflict} &
{} \\
\hline
\hline
%pp3, pp4, pp7 merged
{\cite{9658041,wang2023cooperative,mir2022relay}} & SU/MU (large) MIMO &
\checkmark & \checkmark & \checkmark &  &  &  &
Full-CSI optimization: joint altitude control with weighted minimum mean squared error-based AO, AO with residual error minimization, and coordinated descent over coupled precoder problems \\
\hline
%pp6, pp8, pp5 merged
{\cite{gao2022data,wang2020beamforming,8334304}} & SU/MU massive MISO &
\checkmark & \checkmark & \checkmark & \checkmark & \checkmark &  &
Data-driven HBF with implicit CSI, UAV mobility characterization on coverage, EE, and SE, and channel tracking that integrates flight-control information \\
\hline
%tracking block merged
{\cite{8811740,hyun2021adaptive}} & SU massive MISO, \blue{V2I} &
\checkmark &  & \checkmark & \checkmark & \checkmark & &
Adaptive beam training with angular velocity estimation, and extended Kalman filtering-based tracking with reduced feedback overhead\\
\hline
%new
\blue{\cite{11481114,albataineh2025adaptive,11422871}} & \blue{RIS-assisted / massive SU} &
 &  & \checkmark & \checkmark & \checkmark &  &
\blue{Soft-decision beam training, adaptive beam pairing with local interpolation, and parametric a priori-aided beamspace tracking of the dominant taps}\\
\hline
%limited-feedback block merged
{\cite{shi2018uabeam,shi2021channel,rajashekar2018beamforming,hanna2022distributed}} & SU MISO / MIMO / SISO &
\checkmark &  &  & \checkmark & \checkmark &  &
In-field characterization of feedback overhead and update rate, hybrid feedback schemes, efficient quantization with full diversity, and Kalman filtering against synchronization errors\\
\hline
%greedy block merged
{\cite{10199880,10480178}} & MU MISO / massive MIMO &
 &  & \checkmark & \checkmark & \checkmark &  &
Greedy joint UE and beam selection for weighted sum-SE, and interference-aware greedy beam selection balancing SE and complexity \\
\hline
\hline

%Ours
{Ours} & MU large MISO &
\checkmark & \checkmark & \checkmark & \checkmark & \checkmark & \checkmark  &
Proposed MLR-TG for dynamic switching among GS, sector top-2, global top-2, and global top-3 refinement to balance SE, outage robustness, fairness, and computational complexity\\
\hline

\hline
\end{tabular}
}
% \vspace{-0.25cm}
\end{table*}

The novelty of this work compared with the existing literature is summarized in Table~\ref{tab:novelty_comparison}, and the main contributions are summarized as follows:
\begin{enumerate}
    \item \textbf{\blue{Beam Selection Under Partial Observation}}:
    \blue{We formulate multi-level refinement as the maximization of the net sum-SE conditioned on the information the UAV actually holds at decision time, namely a quantized compact report and its own mobility state. A feedback-only predicted utility assembles a predicted effective channel from the reported complex beam coefficients, discounts it by the aging correlation, and charges each level its own payload fraction. The transmission regularized zero-forcing (RZF) precoder is computed once, from pilot-estimated effective CSI, after the analog subset has been committed, so the analog-selection stage does not require instantaneous transmission-time CSI.}

    \item \textbf{\blue{Multi-Level Policy With a Calibrated Decision Rule:}}
    \blue{We propose the MLR-TG policy, which switches among GS beamforming, sector top-2, global top-2, and global top-3 refinement according to a size-invariant conflict score and an aging-severity metric. The rule is constructed as a calibrated two-statistic approximation motivated by the predicted-utility maximization, its thresholds are fitted offline on a training set disjoint in seed from the evaluation set, and its monotonicity and exact reduction properties are proved. The candidate pool is generated by a bounded-deviation enumeration with a hard cap, which keeps the search tractable as the load grows.}

    \item \textbf{\blue{Causal, Overhead-Accounted Evaluation:}}
    \blue{We evaluate the framework on a three-dimensional (3-D) air-to-ground model with UAV altitude, trajectory, attitude jitter, path-dependent Doppler, and hotspot UE deployment, driving every scheme with one common channel trajectory and evaluating it at the physical instant at which it transmits. Beam sweeping, reporting, decision latency, and pilot overhead are charged to the reported spectral efficiency, the feedback of every scheme is counted in bits, and two oracle bounds separate the loss due to level selection from the loss due to candidate restriction. The system-level simulations (SLSs) report net SE, outage probability (OP), tail rate, fairness, level statistics, threshold sensitivity, and measured complexity.}
\end{enumerate}

\vspace{-0.25cm}
\subsection{Organization and Notational Definitions}

The remainder of this paper is organized as follows. 
Section~\ref{sec:system_model} describes the considered system model, \blue{the 3-D air-to-ground channel model, the causal frame timeline}, compact beam-training feedback, GS baseline, and multi-level candidate refinement.
Section~\ref{sec:problem_formulation} formulates the partially observed selection problem, the feedback-only predicted utility, the conflict metric, the aging-severity metric, and the resulting adaptive multi-level refinement problems. Section~\ref{sec:tg3tr_revisited} presents the TG3-TR strategy and its limitations.
Section~\ref{sec:proposed_method} develops the proposed MLR-TG policy. 
Section~\ref{sec:simulation_results} evaluates performance via extensive simulations. 
Finally, Section~\ref{sec:conclusion} provides concluding remarks of the paper.

\textit{Mathematical Notations:}
$\mathcal{CN}({\mu},\sigma^2)$ and $\mathcal{U}(\cdot)$ denote the complex Gaussian distribution with mean ${\mu}$ and variance $\sigma^2$, and the uniform distribution, respectively. 
$\mathbb{E}\{\cdot\}$ and $\Pr(\cdot)$ denote expectation and probability, respectively. 
$\mathbb{C}^{M\times N}$ denotes the space of $M\times N$ complex-valued matrices, and $\mathbf{I}_N$ denotes the $N\times N$ identity matrix. 
For matrix $\mathbf{A}$, $[\mathbf{A}]_{m,n}$ denotes its $(m,n)$th element. 
The magnitude of scalar $a$ and the cardinality of set $\cal A$ are denoted by $|a|$ and $|\cal A|$, respectively. 
The $\ell_2$-norm of vector $\mathbf{a}$ and the Frobenius norm of matrix $\mathbf{A}$ are denoted by $\|\mathbf{a}\|_2$ and $\|\mathbf{A}\|_F$, respectively. 
$(\cdot)^T$ and $(\cdot)^H$ denote transpose and Hermitian transpose, respectively. 
% $\mathbf{A}^H$ also represents the conjugate transpose of $\mathbf{A}$. 
$\lceil \cdot \rceil$ denotes the ceiling (upper bound) operator.
$\mathbb{1}(\cdot)$ denotes the indicator function. 
For set $\mathcal{A}$ and element $a\in\mathcal{A}$, $\mathcal{A}\!\setminus\!\{ a \}$ denotes the set excluding $a$. 
Unless otherwise specified, uppercase boldface letters denote matrices, lowercase boldface letters denote vectors, scalar quantities are denoted by lowercase letters, and calligraphic letters denote sets.
Furthermore, $k \in {\cal K}$ is assumed throughout this paper.

% \textcolor{red}{Abbreviations not defined:  SNR, SINR, AG3}

\section{System Model}
\label{sec:system_model}

% \textcolor{red}{Start to check from this section.}

\subsection{Multi-User HBF mmWave UAV Downlink}
\label{subsec:downlink_model}

\begin{figure} [!t]
	\centering
	%	\captionsetup{justification=centering}
	\includegraphics[width=\linewidth, angle=0]{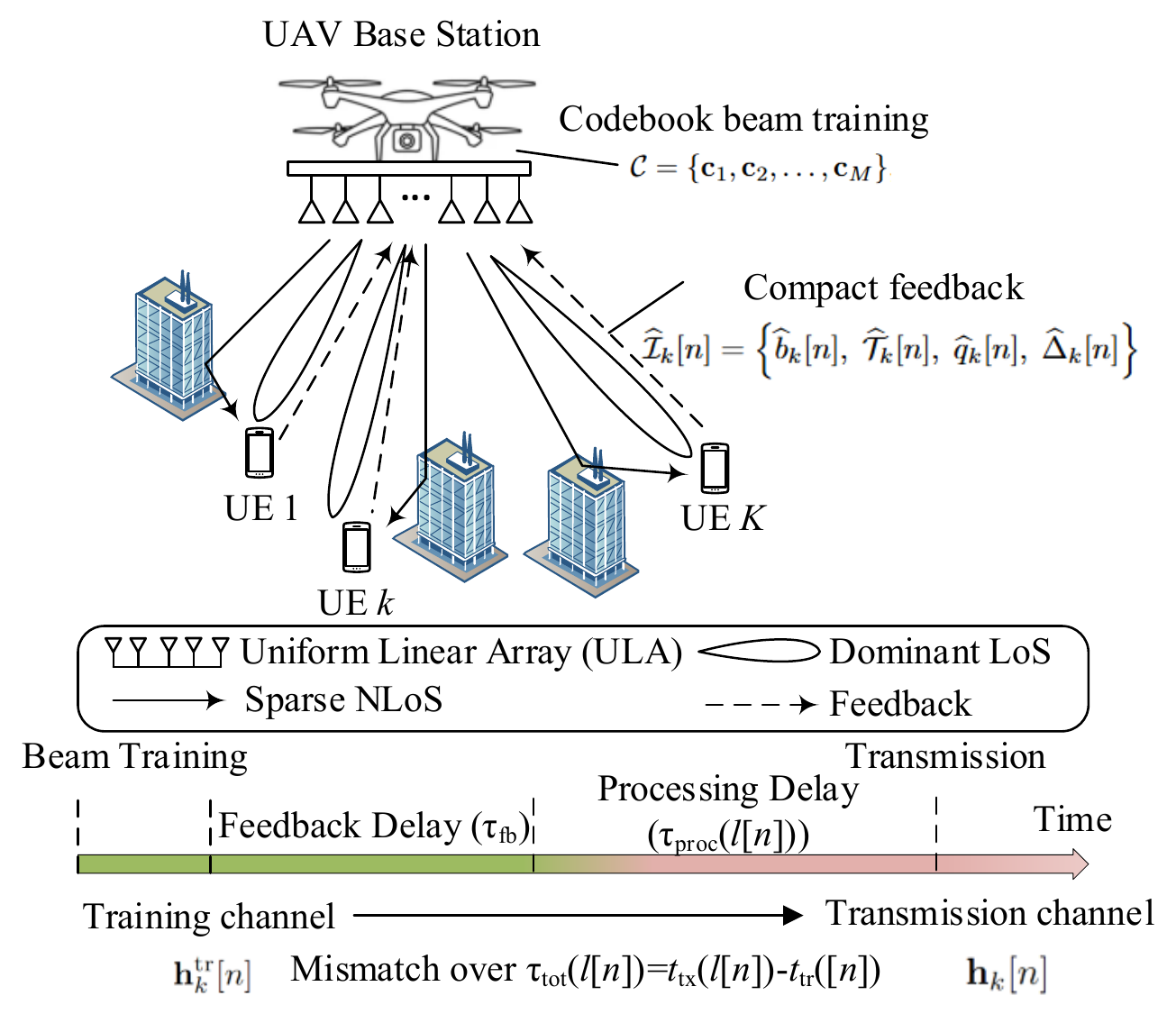}
	\caption{System model of the considered MU mmWave UAV downlink with codebook-based beam training and compact beam feedback. }
   \label{fig:system_model}
   % \vspace{-0.25cm}
\end{figure}

We consider an MU mmWave UAV downlink system, as illustrated in Fig.~\ref{fig:system_model}, where a UAV equipped with a uniform linear array (ULA) of $N_t$ antennas and $N_{\mathrm{RF}}$ RF chains serves $K$ single-antenna ground UEs \cite{9658041,wang2020beamforming}. 
Let $\mathcal{K}=\{1,\ldots,K\}$ denote the set of $K$ UEs. 
To balance beamforming capability and hardware cost, the UAV adopts a hybrid analog-digital precoding architecture, in which the analog stage provides directional beamforming gain, while the digital stage mitigates inter-user interference over the reduced-dimensional effective channel.

% The training-to-transmission interval includes beam-sweep/reporting time, feedback latency, refinement-dependent processing latency, and effective-channel acquisition, resulting in the total mismatch $\tau_{\mathrm{tot}}(\ell[n])$ defined in \eqref{eq:total_delay}.

At beam-management frame $n$, a digital baseband precoder $\mathbf{F}_{\mathrm{BB}}[n]
= [\mathbf{f}_1[n],...,\mathbf{f}_k[n],...,\mathbf{f}_K[n]]
\in\mathbb{C}^{N_{\mathrm{RF}}\times K}$ is employed, where 
$\mathbf{f}_k[n] \in \mathbb{C}^{N_{\mathrm{RF}}\times 1}$ denotes the baseband precoding vector for UE $k$.\footnote{Unless stated otherwise, $[n]$ indexes the beam-management frame throughout this paper. \blue{A frame has duration $T_{\mathrm{f}}$, spans several transmission slots, and carries one beam-training sweep and one beam decision, so all quantities written with $[n]$ are constant within it.}}
Subsequently, an analog RF precoder $\mathbf{F}_{\mathrm{RF}}[n]\in\mathbb{C}^{N_t\times N_{\mathrm{RF}}}$ is constructed such that
all entries of $\mathbf{F}_{\mathrm{RF}}[n]$ have constant modulus,
i.e.,
$|[\mathbf{F}_{\mathrm{RF}}[n]]_{p,q}| = 1/\sqrt{N_t}, \forall p,q$.
Accordingly, the transmitted signal at the UAV is given by
% At transmission interval $n$, the UAV transmits $K$ independent data streams collected in $\mathbf{s}[n]\in\mathbb{C}^{K\times 1}$ with $\mathbb{E}\!\left[\mathbf{s}[n]\mathbf{s}^H[n]\right]=\mathbf{I}_K$. 
% The transmitted signal is given by
\begin{equation}
{\bf{x}}[n] = {{\bf{F}}_{{\rm{RF}}}}[n]{{\bf{F}}_{{\rm{BB}}}}[n]{\bf{s}}[n] = \sum\nolimits_{k \in {\cal K}} {{{\bf{F}}_{{\rm{RF}}}}[n]{{\bf{f}}_k}[n]{s_k}{{[n]}}} ,
\label{eq:tx_signal}
\end{equation}
% where $\mathbf{F}_{\mathrm{RF}}[n]\in\mathbb{C}^{N_t\times N_{\mathrm{RF}}}$ and $\mathbf{F}_{\mathrm{BB}}[n]\in\mathbb{C}^{N_{\mathrm{RF}}\times K}$ denote the analog and digital precoders, respectively. 
where $\mathbf{s}[n]
=[s_1[n],...,s_k[n],...,s_K[n]]
\in\mathbb{C}^{K\times 1}$ is the vector comprising transmitted symbols of all UEs, satisfying 
$\mathbb{E}\big\{ {{{\left| {{s_k}{{[n]}}} \right|}^2}} \big\} = 1,\forall k$.
% $\mathbb{E}\!\left[\mathbf{s}[n]\mathbf{s}^H[n]\right]=\mathbf{I}_K$. 
Under the practical transmit power constraint at the UAV, $\mathbf{F}_{\mathrm{BB}}[n]$ is normalized to satisfy
% The overall hybrid precoder is subject to the transmit power constraint
\cite{tu2025semi,tu2026hybrid,tu2026multi}
\begin{equation}
% \left\|\mathbf{F}_{\mathrm{RF}}[n]\mathbf{F}_{\mathrm{BB}}[n]\right\|_F^2 \le P_t.
\sum\nolimits_{k \in {\cal K}} {\left\| {{{\bf{F}}_{{\rm{RF}}}}[n]{{\bf{f}}_k}[n]} \right\|_2^2}  \le {P_t}.
\label{eq:power_constraint}
\end{equation}
Consequently, the received signal at UE $k$ is expressed as
\begin{equation}
{y_k}[n] = {\bf{h}}_k^H[n]{{\bf{F}}_{{\rm{RF}}}}[n]\!\!\sum\nolimits_{j \in {\cal K}}\!\!{{\bf{f}}_j}[n]{s_j}{{[n]}}  + {z_k}[n],
\label{eq:rx_signal}
\end{equation}
where $\mathbf{h}_k[n]\in\mathbb{C}^{N_t\times 1}$ denotes the channel coefficient vector between the UAV and UE $k$,
while $z_k[n]\sim\mathcal{CN}(0,\sigma^2)$ is the zero-mean additive Gaussian white noise at UE $k$ with variance $\sigma^2$.
Based on \eqref{eq:rx_signal}, the instantaneous received signal-to-interference-plus-noise ratio (SINR) of UE $k$ is computed by
\begin{equation}
% \gamma_k[n]
% =
% \frac{\rho \left|\mathbf{h}_k^H[n]\mathbf{f}_k[n]\right|^2}
% {1+\rho \sum_{j\neq k}\left|\mathbf{h}_k^H[n]\mathbf{f}_j[n]\right|^2},
{\gamma _k}[n] = \frac{{{{\left| {{\bf{h}}_k^H[n]{{\bf{F}}_{{\rm{RF}}}}[n]{{\bf{f}}_k}[n]} \right|}^2}}}{{\sum\nolimits_{j \in {\cal K}\backslash \{k\}} {{{\left| {{\bf{h}}_k^H[n]{{\bf{F}}_{{\rm{RF}}}}[n]{{\bf{f}}_j}[n]} \right|}^2}}  + {\sigma ^2}}}.
\label{eq:sinr_user}
\end{equation}
% \textcolor{red}{Check until this...}
% where $\rho=P_t/\sigma^2$ denotes the transmit SNR. 
Accordingly, the achievable sum-SE is expressed as
\begin{equation}
R_{\mathrm{sum}}[n]
=
\sum\nolimits_{k\in {\cal S}[n]}\log_2\!\left(1+\gamma_k[n]\right),
\label{eq:sum_rate}
\end{equation}
\blue{where ${\cal S}[n]\subseteq{\cal K}$ with $|{\cal S}[n]|=\min(K,N_{\mathrm{RF}})$ is the set of UEs scheduled in frame $n$, which is defined in Section~\ref{subsec:gs_and_multilevel_refinement}. When $K\le N_{\mathrm{RF}}$ every UE is served and ${\cal S}[n]={\cal K}$.}

% \textcolor{red}{Check until this...}

\subsection{Three-Dimensional Air-to-Ground Channel with Platform Attitude Dynamics}
\label{subsec:channel_model}

The UAV--UE links are modeled as sparse mmWave channels with a limited number of dominant propagation paths.
This model is well suited for UAV-assisted mmWave systems because aerial links are often LoS-dominant and highly directional, while platform mobility makes beam alignment and channel tracking particularly sensitive to user geometry and motion \cite{8288376,wang2020beamforming}.
% 10212563,9909004
%
{\color{black}
Rather than drawing path angles independently at each frame, we generate one continuous-time channel trajectory per topology from the 3-D geometry, so that angular drift, path-dependent Doppler, and path loss all follow from the motion of the platform \cite{3gpp36777,6863654}.
The UAV flies at altitude $H_{\mathrm{U}}$ with the constant velocity vector $\mathbf{v}=v[\cos\psi_{\mathrm{v}},\sin\psi_{\mathrm{v}},0]^T$, so its position at continuous time $t$ is $\mathbf{p}_{\mathrm{U}}(t)=\mathbf{p}_{\mathrm{U}}(0)+\mathbf{v}t$.
The $K$ UEs are dropped on the ground inside an annulus of radii $r_{\min}$ and $R_{\mathrm{c}}$, a fraction $\varrho_{\mathrm{h}}$ of them inside $N_{\mathrm{h}}$ hotspots of radius $r_{\mathrm{h}}$, because co-angular UEs are precisely the case in which beam conflict occurs, and the $l$-th path of UE $k$ departs towards the scatterer position $\mathbf{q}_{k,l}$, with $\mathbf{q}_{k,1}$ the UE itself and $\mathbf{q}_{k,l}$, $l\ge2$, drawn in a ground-clutter ring around it.
Since a ULA resolves only the projection of a departure direction onto its axis, writing ${\mathbf{u}_{\mathrm{a}}(t)}$ for the unit vector of that axis, the direction cosine of path $(k,l)$ is expressed as
\begin{equation}
\Omega_{k,l}(t)
=
{\mathbf{u}^T_{\mathrm{a}}(t)}\frac{\mathbf{q}_{k,l}-\mathbf{p}_{\mathrm{U}}(t)}{d_{k,l}(t)},
\
d_{k,l}(t)=\big\|\mathbf{q}_{k,l}-\mathbf{p}_{\mathrm{U}}(t)\big\|_2 ,
\label{eq:dircos}
\end{equation}
and the corresponding normalized array response vector is written as
\begin{equation}
\mathbf{a}(\Omega)
=
\big[
1,
e^{j\pi\Omega},
\dots,
e^{j\pi (N_t-1)\Omega}
\big]^T/\sqrt{N_t},
\label{eq:array_response_journal_new}
\end{equation}
for inter-element spacing $d=\lambda/2$, where $\lambda$ is the carrier wavelength.
The classical form $\Omega=\sin\theta$ is recovered when the departure direction lies in the plane containing the array axis.
A rotary-wing UAV holding a track continuously corrects its attitude, and the resulting angular rate is far larger than the rate at which translation changes the departure angles.
The array axis therefore carries a yaw--pitch jitter, i.e., ${\mathbf{u}_{\mathrm{a}}(t)}=[\cos\psi(t)\cos\chi(t),\sin\psi(t)\cos\chi(t),\sin\chi(t)]^T$ with
\begin{equation}
\psi(t)=\psi_{\mathrm{v}}+\frac{\Omega_{\mathrm{y}}}{2\pi f_{\mathrm{a}}}
\big[\sin(2\pi f_{\mathrm{a}}t+\phi_{\mathrm{y}})-\sin\phi_{\mathrm{y}}\big],
\label{eq:attitude}
\end{equation}
and an analogous expression for the pitch $\chi(t)$, where $f_{\mathrm{a}}$ is the attitude-loop frequency and the peak rate $\Omega_{\mathrm{y}}=\omega_0+\omega_1 v/v_{\mathrm{ref}}$ grows with airspeed.
The channel of UE $k$ at time $t$ is then written as
\begin{align}
\mathbf{h}_k(t)
=
\sqrt{G_k}\sum\nolimits_{l=1}^{L}
&\sqrt{p_{k,l}}\,g_{k,l}(t)\,
e^{-j2\pi d_{k,l}(t)/\lambda + j\varphi_{k,l}}\nonumber\\
&\times\mathbf{a}\big(\Omega_{k,l}(t)\big),
\label{eq:mmwave_channel_training}
\end{align}
where $L$ is the number of dominant paths, $G_k$ is the large-scale gain obtained from the free-space path loss, the LoS probability of \cite{6863654} with its associated non-line-of-sight (NLoS) excess loss, and log-normal shadowing, while $p_{k,1}=\kappa_{\mathrm{R}}/(\kappa_{\mathrm{R}}+1)$ and $p_{k,l}=1/[(\kappa_{\mathrm{R}}+1)(L-1)]$, $l\ge2$, split the power according to the Rician factor $\kappa_{\mathrm{R}}$, and $\varphi_{k,l}$ is a random initial phase.
The path-dependent Doppler shift needs no separate parameter, since it is the time derivative of the propagation phase in \eqref{eq:mmwave_channel_training}, i.e., $f_{k,l}^{\mathrm{D}}(t)=-\lambda^{-1}\mathbf{v}^T(\mathbf{q}_{k,l}-\mathbf{p}_{\mathrm{U}}(t))/d_{k,l}(t)$.
The dominant component is specular, i.e., $g_{k,1}(t)=1$ for a LoS UE, whereas each scattered component follows Clarke's sum-of-sinusoids process
$g_{k,l}(t)=N_{\mathrm{o}}^{-1/2}\sum\nolimits_{i=1}^{N_{\mathrm{o}}}e^{j(2\pi f_{\mathrm{D}}^{\max}t\cos\alpha_i+\phi_i)}$
with $f_{\mathrm{D}}^{\max}=v/\lambda$ and $\alpha_i,\phi_i\sim{\cal U}(0,2\pi)$ drawn once per topology, so that the same realization is reused by every scheme.

\blue{Beam training and data transmission occur at two different physical instants, and the interval between them is what makes the beam decision outdated.
Let $t_{\mathrm{tr}}[n]$ denote the instant at which the beam sweep of frame $n$ is measured and $t_{\mathrm{tx}}(\ell[n])$ the instant at which the payload of frame $n$ is transmitted under refinement level $\ell[n]\in\{0,1,2,3\}$, where $\ell[n]=0$ corresponds to the GS baseline, $\ell[n]=1$ to sector top-$2$ refinement, $\ell[n]=2$ to global top-$2$ refinement, and $\ell[n]=3$ to global top-$3$ refinement, as detailed in Section~\ref{subsec:gs_and_multilevel_refinement}.
Both instants are read from the frame timeline of Section~\ref{subsec:frame_timeline}.
The training and transmission channels of UE $k$ are therefore $\mathbf{h}_k^{\mathrm{tr}}[n]=\mathbf{h}_k(t_{\mathrm{tr}}[n])$ and $\mathbf{h}_k[n]=\mathbf{h}_k(t_{\mathrm{tx}}(\ell[n]))$, and the mismatch between them is governed by
}
\begin{equation}
\tau_{\mathrm{tot}}(\ell[n])
=
t_{\mathrm{tx}}(\ell[n]) - t_{\mathrm{tr}}[n]
=
\tau_{\mathrm{sw}}+\tau_{\mathrm{fb}}+\tau_{\mathrm{proc}}(\ell[n])+\tau_{\mathrm{p}},
\label{eq:total_delay}
\end{equation}
where $\tau_{\mathrm{sw}}$ collects the second half of the beam sweep and the uplink report, $\tau_{\mathrm{fb}}$ is the feedback latency, $\tau_{\mathrm{proc}}(\ell[n])$ is the decision latency of level $\ell[n]$, and $\tau_{\mathrm{p}}$ is the duration of the pilots used for effective-channel acquisition.}

\blue{For analytical tractability, the pair $(\mathbf{h}_k^{\mathrm{tr}}[n],\mathbf{h}_k[n])$ is summarized by the first-order Gauss--Markov abstraction}
\begin{equation}
\mathbf{h}_k[n]
=
\rho(\ell,n)\,\mathbf{h}_k^{\mathrm{tr}}[n]
+
\sqrt{1-\rho^2(\ell,n)}\,\mathbf{e}_k[n],
\label{eq:aged_channel_model}
\end{equation}
where $\mathbf{e}_k[n]$ denotes the innovation component, \blue{which is uncorrelated with $\mathbf{h}_k^{\mathrm{tr}}[n]$ and carries the same second-order statistics}, and $\rho(\ell,n)\in[0,1]$ is the temporal correlation coefficient. Following a standard exponential aging model,
\begin{equation}
\rho(\ell,n)
=
\exp\bigl(-\beta f_D[n]\tau_{\mathrm{tot}}(\ell[n])\bigr),
\label{eq:rho_model}
\end{equation}
where $\beta>0$ is a correlation constant and $f_D[n]={v[n]}/{\lambda}$ is the Doppler frequency associated with the UAV speed $v[n]$ \cite{10770152,8811740}. Therefore, a higher UAV speed or a longer delay reduces channel correlation and increases the risk that a beam selected based on training information becomes outdated before data transmission.

\begin{remark}[\blue{Scope of the Aging Abstraction}]
\blue{Equation \eqref{eq:rho_model} enters only where a closed-form predictor is required, namely the candidate metric of Section~\ref{subsec:predicted_utility}.
Every reported result is generated with the geometry-based model \eqref{eq:mmwave_channel_training}, so beam misalignment caused by angular drift and attitude jitter is reproduced explicitly instead of being absorbed into a scalar.
A per-UE common phase rotation leaves the achievable SINRs unchanged, so the quantity that \eqref{eq:rho_model} has to reproduce is the phase-invariant correlation of the reported beamspace signature rather than that of the raw channel vector.
Section~\ref{sec:simulation_results} fits $\beta$ to the geometry-based model and reports the residual error of the fit.}
\end{remark}

\vspace{-0.25cm}
%----------------
\subsection{\blue{Causal Frame Timeline and Net Spectral Efficiency}}
\label{subsec:frame_timeline}

{\color{black}
Beam management consumes air time, and a deeper candidate search consumes more of it.
A comparison that charges every scheme the same overhead would therefore hide the very trade-off this paper is about.
Each beam-management frame of duration $T_{\mathrm{f}}$ is accordingly split into an exhaustive downlink beam sweep of $M$ symbols, an uplink report of $K$ symbols, the feedback latency $\tau_{\mathrm{fb}}$, the decision latency $\tau_{\mathrm{proc}}(\ell[n])$, $N_{\mathrm{p}}$ demodulation pilot symbols used to acquire the effective channel, and the remaining payload.
With $T_{\mathrm{s}}$ the symbol duration, the decision of frame $n$ becomes available at
\begin{equation}
t_{\mathrm{tx}}(\ell)=\big(M+K+N_{\mathrm{p}}\big)T_{\mathrm{s}}+\tau_{\mathrm{fb}}+\tau_{\mathrm{proc}}(\ell),
\label{eq:tdec}
\end{equation}
measured from the start of the frame.
A scheme whose decision is not ready before the end of the current frame is pipelined to the next frame boundary, so the payload fraction is calculated as
\begin{equation}
\eta(\ell)=\frac{N_{\mathrm{w}}(\ell)T_{\mathrm{f}}-t_{\mathrm{tx}}(\ell)}{N_{\mathrm{w}}(\ell)T_{\mathrm{f}}},
\
N_{\mathrm{w}}(\ell)=\Big\lceil \frac{t_{\mathrm{tx}}(\ell)}{T_{\mathrm{f}}}\Big\rceil .
\label{eq:payload_fraction}
\end{equation}
The figure of merit used throughout this paper is therefore the net, or effective, sum-SE
\begin{equation}
\bar R_{\mathrm{sum}}\big({\cal B}[n],\ell[n];n\big)
=
\eta(\ell[n])\,R_{\mathrm{sum}}\big({\cal B}[n],\ell[n];n\big),
\label{eq:net_se}
\end{equation}
in which $R_{\mathrm{sum}}$ of \eqref{eq:sum_rate} is evaluated on the channel realization at the physical instant $t_{\mathrm{tx}}(\ell[n])$ at which the scheme actually transmits.
Every scheme is driven by the same channel trajectory and the same random draws, and differs only through the instant at which it transmits and the payload it retains.
With the parameters of Table~\ref{tab:simulation_parameters}, the four levels transmit at $t_{\mathrm{tx}}=2.35$, $2.48$, $2.54$, and $2.84$~ms after the start of the frame and retain payload fractions of $\eta(0\!:\!3)=(0.765,0.753,0.746,0.716)$.
}

\vspace{-0.25cm}
%----------------
\subsection{Codebook-Based Beam Training and Compact Feedback}
\label{subsec:beam_training_feedback_new}

% It is worth mentioning that selecting the analog beam subset is challenging due to limited beam-training feedback under UAV mobility, while the digital precoder is subsequently designed based on the resulting effective channel.
% Therefore, the quality of analog beam selection directly affects not only the achievable beamforming gain but also the interference-suppression capability of the digital stage.
The quality of analog beam selection under limited beam-training feedback affects not only the beamforming gain but also the interference-suppression capability of the subsequent digital stage.
The UAV adopts a predefined beam codebook
\begin{equation}
\mathcal{C}=\{\mathbf{c}_1,\dots,\mathbf{c}_m,\dots,\mathbf{c}_M\},
\label{eq:beam_codebook_journal_new}
\end{equation}
where each codeword $\mathbf{c}_m\in\mathbb{C}^{N_t\times 1}$ is a constant-modulus beamforming vector derived based on the ULA structure and $M$ is the beam codebook dimension. 
% For $S<2$, the normalized sector- and beam-collision terms are defined as zero. 
It is worth mentioning that codebook-based beam training is widely used in mmWave systems as a practical alternative to full CSI acquisition, especially under limited feedback and mobility constraints \cite{6717211,7389996}. The codebook is partitioned into $N_s$ disjoint angular sectors such that
\begin{equation}
\mathcal{C}
=
\bigcup\limits_{q=1}^{N_s}\mathcal{C}_q,
\,\,
\mathcal{C}_q\cap\mathcal{C}_{q'}=\emptyset,
\,\, \forall q'\neq q, 
\,\, q' \in \{1,\dots,N_s\}.
\label{eq:sector_partition_journal_new}
\end{equation}

During beam training, UE $k$ measures the beam gains over the training channel as
\begin{equation}
g_{k,m}^{\mathrm{tr}}[n]
=
\big|
\left(\mathbf{h}_k^{\mathrm{tr}}[n]\right)^H\mathbf{c}_m
\big|^2,
\ \forall m \in \{1,\dots,M\}.
\label{eq:beam_gain_journal_new}
\end{equation}
Instead of feeding back full CSI, each UE returns a compact feedback packet composed of beam-training indicators.
Such compact reporting is more suitable for mobility-sensitive mmWave UAV systems, where frequent high-dimensional CSI updates would incur substantial overhead \cite{8288376,8395281,albataineh2025adaptive}.
{\color{black}
The packet has to be rich enough for every refinement level to be executable from it alone.
UE $k$ therefore ranks the codebook by \eqref{eq:beam_gain_journal_new}, keeps its $L_{\mathrm{r}}$ strongest codewords, and reports their indices together with the corresponding quantized complex coefficients, i.e.,
\begin{equation}
\widehat{\mathcal{I}}_k[n]
=
\Big\{
\widehat{q}_k[n],\;
\big(\widehat{b}_{k,i}[n],\,\widehat{c}_{k,i}[n]\big)_{i=1}^{L_{\mathrm{r}}}
\Big\},
\label{eq:compact_feedback_packet}
\end{equation}
where $\widehat{q}_k[n]$ is the dominant sector index, $\widehat{b}_{k,1}[n],\ldots,\widehat{b}_{k,L_{\mathrm{r}}}[n]$ are the reported codeword indices ordered by decreasing gain, and $\widehat{c}_{k,i}[n]$ is the quantized version of $(\mathbf{h}_k^{\mathrm{tr}}[n])^H\mathbf{c}_{\widehat{b}_{k,i}[n]}$, obtained with $q_{\mathrm{A}}$ bits for its amplitude in the logarithmic domain and $q_{\mathrm{P}}$ bits for its phase.}
\blue{The reported candidate list and the quantized beam-gap indicator used by the conflict metric of Section~\ref{subsec:conflict_metric_journal} are read from the same packet, namely $\widehat{\mathcal{T}}_k[n]=\{\widehat{b}_{k,i}[n]\}_{i=1}^{L_{\mathrm{r}}}$ and $\widehat{\Delta}_k[n]=20\log_{10}(|\widehat{c}_{k,1}[n]|/|\widehat{c}_{k,2}[n]|)$, so they cost no extra bits.
Reporting complex coefficients instead of gains is what makes the framework self-consistent: the UAV can score any candidate subset drawn from reported codewords without acquiring any further channel knowledge, which is examined in Section~\ref{subsec:predicted_utility}.
%The resulting uplink payload is $B_{\mathrm{UE}}=L_{\mathrm{r}}\lceil\log_2 M\rceil+\lceil\log_2 N_s\rceil+L_{\mathrm{r}}(q_{\mathrm{A}}+q_{\mathrm{P}})$ bits per UE per frame, against $N_t(q_{\mathrm{A}}+q_{\mathrm{P}})$ bits for full-CSI reporting and $Mq_{\mathrm{A}}$ bits for reporting one quality metric per codeword.
The resulting uplink payload is $B_{\mathrm{UE}}=L_{\mathrm{r}}\lceil\log_2 M\rceil+\lceil\log_2 N_s\rceil+L_{\mathrm{r}}(q_{\mathrm{A}}+q_{\mathrm{P}})$ bits per UE per frame. For comparison, full-CSI reporting requires $N_t(q_{\mathrm{A}}+q_{\mathrm{P}})$ bits per UE, while a full beam-quality report with one $q_g=6$-bit scalar metric per codeword requires $Mq_g$ bits per UE.
Table~\ref{tab:overhead} in Section~\ref{sec:simulation_results} evaluates all of these for the considered configuration.
}

\subsubsection*{\blue{Effective-channel acquisition}}
{\color{black}
The reported coefficients are aged by $\tau_{\mathrm{tot}}(\ell[n])$ and coarsely quantized, which is tolerable for ranking candidate subsets but not for inverting an $N_{\mathrm{RF}}\times N_{\mathrm{RF}}$ effective channel.
The digital precoder is therefore built only after the analog subset has been fixed.
Once $\mathbf{F}_{\mathrm{RF}}[n]$ is applied, $N_{\mathrm{p}}$ dedicated pilot symbols are transmitted through it and each scheduled UE estimates and reports its $N_{\mathrm{RF}}$-dimensional effective channel with $q_{\mathrm{eff}}$ bits per coefficient, so that the UAV obtains
\begin{equation}
\widehat{\mathbf{H}}_{\mathrm{eff}}[n]=\mathbf{H}_{\mathrm{eff}}[n]+\mathbf{E}[n],
\quad
[\mathbf{E}[n]]_{k,i}\sim{\cal CN}(0,\sigma_{\mathrm{e}}^2),
\label{eq:heff_estimate}
\end{equation}
with $\sigma_{\mathrm{e}}^2=\sigma^2/(\rho_{\mathrm{p}}N_{\mathrm{p}})$ and $\rho_{\mathrm{p}}$ the pilot-power boost.
This stage is common to every scheme compared in Section~\ref{sec:simulation_results}, is charged to all of them in \eqref{eq:tdec}, and is the only place where instantaneous channel knowledge enters the proposed framework.
}

\vspace{-0.25cm}
%----------------
\subsection{GS Baseline and Multi-Level Candidate Refinement}
\label{subsec:gs_and_multilevel_refinement}

{\color{black}
Only $N_{\mathrm{RF}}$ data streams can be supported per frame, so a scheduled set has to be defined whenever $K>N_{\mathrm{RF}}$.
Let $\bar{R}_k[n]$ denote the exponentially averaged net rate delivered to UE $k$ and let $\mathcal{S}[n]\subseteq\mathcal{K}$ with $|\mathcal{S}[n]|=\min(K,N_{\mathrm{RF}})$ denote the scheduled set.
UEs are admitted in decreasing order of the proportional-fair metric $\log_2(1+|\widehat{c}_{k,1}[n]|^2)/\bar{R}_k[n]$, an admitted UE whose strongest codeword is already taken being moved to the next entry of its reported list, and $\bar{R}_k[n]$ being updated after each frame with forgetting factor $\epsilon_{\mathrm{pf}}$.
Fairness is therefore accounted over time rather than within a frame, and unscheduled UEs contribute a zero rate to the averages of Section~\ref{sec:simulation_results}.
The beam-selection problem below is stated for $\mathcal{S}[n]$, and all sums over $\mathcal{K}$ in \eqref{eq:sum_rate} and \eqref{eq:sinr_user} are restricted to $\mathcal{S}[n]$.}
\blue{For an unscheduled UE $k\notin\mathcal{S}[n]$, we set $\mathbf{f}_k[n]=\mathbf{0}$. Hence, the signal and power expressions remain valid, while only the scheduled UEs contribute nonzero transmitted streams.}

Using the compact feedback, the UAV first constructs a low-complexity GS baseline that is referred to as $\ell[n] =0$.
The strongest reported beam of each \blue{scheduled} UE is treated as its preferred beam, duplicate beams are removed, and any missing RF chains are completed using strong unused beams from the reported candidates \cite[Alg.~1]{wu2023simultaneous}. Let
\begin{align}
\label{eq:GS_baseline}
    \mathcal{B}_{\mathrm{GS}}[n]
    =
    \{b_1[n],\dots,b_{N_{\mathrm{RF}}}[n]\}
\end{align}
denote the resulting GS beam subset, where $b_i[n] \in \{ 1, \dots,M \}$ is the index of the selected codeword for the $i$-th RF chain.
Accordingly, the analog precoder is constructed as
\begin{equation}
\mathbf{F}_{\mathrm{RF}}^{\mathrm{GS}}[n]
=
\big[
\mathbf{c}_{b_1[n]},\dots,\mathbf{c}_{b_{N_{\mathrm{RF}}}[n]}
\big].
\label{eq:frf_gs_journal_new}
\end{equation}

Given an analog beam subset, the digital precoder is designed using RZF. For the greedy baseline, the effective channel is given as
\begin{equation}
\mathbf{H}_{\mathrm{eff}}^{\mathrm{GS}}[n]
=
(\mathbf{H}[n])^H\mathbf{F}_{\mathrm{RF}}^{\mathrm{GS}}[n],
\label{eq:effective_channel_gs_journal_new}
\end{equation}
where $\mathbf{H}[n]
=
\left[
\mathbf{h}_1[n],\dots,\mathbf{h}_{|{\cal S}[n]|}[n]
\right] \in \mathbb{C}^{N_t \times |{\cal S}[n]|}$ is the concatenated channels \blue{of the scheduled UEs}.
Subsequently, the digital precoder is constructed \blue{from the pilot-based estimate $\widehat{\mathbf{H}}_{\mathrm{eff}}^{\mathrm{GS}}[n]$ of \eqref{eq:heff_estimate}} as
\begin{equation}
\mathbf{F}_{\mathrm{BB}}^{\mathrm{GS}}[n]
=
\big(\widehat{\mathbf{H}}_{\mathrm{eff}}^{\mathrm{GS}}[n]\big)^H
\Big(
\widehat{\mathbf{H}}_{\mathrm{eff}}^{\mathrm{GS}}[n]
\big(\widehat{\mathbf{H}}_{\mathrm{eff}}^{\mathrm{GS}}[n]\big)^H
+\mu\mathbf{I}_{|{\cal S}[n]|}
\Big)^{-1},
\label{eq:bb_gs_journal_new}
\end{equation}
where $\mu \buildrel \Delta \over =|{\cal S}[n]|/\rho_{\mathrm{tx}} >0$ is the regularization parameter and $\rho_{\mathrm{tx}} \buildrel \Delta \over = {P_t}/{\sigma ^2}$ is the transmit signal-to-noise ratio (SNR).
Finally, $\mathbf{F}_{\mathrm{BB}}^{\mathrm{GS}}[n]$ in \eqref{eq:bb_gs_journal_new} is normalized to satisfy \eqref{eq:power_constraint}.

% Although the GS is computationally efficient, it may become suboptimal when the feedback indicators are ambiguous, outdated, or heavily affected by multi-user beam contention. 
% To improve robustness without always invoking a full candidate search, the UAV is allowed to activate one of several refinement levels. 
% %
% Specifically, level $\ell[n]=1$ performs sector top-$2$ refinement, selecting two strongest beams within the reported sector.
% Meanwhile, levels $\ell[n]=2$ and $\ell[n]=3$ perform global top-$2$ and top-$3$ refinements, selecting the two and three strongest beams over the entire codebook, respectively.
% % Furthermore, level $\ell=3$ performs global top-$3$ refinement, selecting the three strongest beams. 
% The selected refinement level thus determines both the dimension of the candidate beam pool and the associated processing delay in \eqref{eq:total_delay}.
% % Accordingly, the key design problem is not only whether to activate refinement, but also how deep the refinement should be under the current contention and channel aging conditions.
% The problem formulation of the multi-level candidate refinement is discussed in Section~\ref{sec:problem_formulation}.
Although the GS is computationally efficient, it may become suboptimal when the feedback indicators are ambiguous, outdated, or heavily affected by MU beam contention. To improve robustness without always invoking a full candidate search, the UAV is allowed to activate one of the refinement levels $\ell[n]\in\{1,2,3\}$ defined in Section~\ref{subsec:channel_model} and constructed in Section~\ref{subsec:level_dependent_pool}, which determines both the candidate pool dimension and the processing delay in \eqref{eq:total_delay}.

\begin{remark}[\blue{Ordering of the Candidate Pools}]
\blue{The pools of different levels are ordered by search cardinality, not by set inclusion.
A sector-local second beam need not belong to the global top-$2$ or top-$3$ list of the same UE, and duplicate removal followed by beam completion can return subsets that no larger-level pool contains.
Levels $1$ and $2$ moreover explore two different regions of the codebook at the same cardinality, one restricted to the reported sector and one spanning all sectors.
No nesting property is therefore claimed, and the benefit of a level comes from where it searches as much as from how widely it searches.}
\end{remark}

%--------------
\section{Problem Formulation}
\label{sec:problem_formulation}

It is worth mentioning that the UAV does not perform analog beam selection from perfect instantaneous CSI at each frame. Instead, it makes beam management decisions based on compact beam-training feedback of
$\widehat{\mathcal{I}}[n] = [\widehat{\mathcal{I}}_1[n],\cdots,\widehat{\mathcal{I}}_k[n],\cdots,\widehat{\mathcal{I}}_K[n]]$ \cite{8458146}.
Based on this limited and delayed feedback, the UAV must determine not only whether candidate refinement should be activated, but also the refinement depth and the final analog beam subset.
% \[
% \widehat{\mathcal{I}}[n]
% =
% \big\{
% \widehat{\mathcal{I}}_k[n]
% \big\}_{k\in {\cal K}},
% \]
% where compact feedback $\widehat{\mathcal{I}}_k[n]$ contains the strongest reported beam, a short candidate beam list, the dominant sector index, and a quantized beam-gap indicator of user $k$. 
% Based on this limited and delayed feedback, the UAV must determine not only whether candidate refinement should be activated, but also the refinement depth and the final analog beam subset.
% The key challenge lies in balancing beam selection accuracy and processing delay under channel aging. While deeper candidate refinement can improve analog beam selection, it incurs additional processing latency, which exacerbates the training-transmission mismatch in UAV mobility.
% The key challenge is that the analog beam subset is selected from compact beam-training feedback under UAV mobility, while the digital precoder is subsequently designed based on the resulting effective channel. Consequently, the quality of analog beam selection affects not only the beamforming gain but also the interference suppression capability of the digital stage. Moreover, deeper candidate refinement can improve beam selection accuracy, but it introduces additional processing delay, thereby exacerbating the training-transmission mismatch caused by channel aging.

%-------
\subsection{Joint Refinement-Level and Beam-Subset Selection}
\label{subsec:joint_problem_journal}

Given the compact feedback $\widehat{\mathcal{I}}[n]$, the UAV jointly determines refinement level $\ell[n]$ and analog beam subset $\mathcal{B}[n]$ to maximize the achievable sum-SE during data transmission.
\blue{The transmission-time channel $\mathbf{H}[n]$ is not observable when the decision is taken, so the objective is the net sum-SE of \eqref{eq:net_se} conditioned on what the UAV actually knows}, namely the compact report and its own mobility state $(v[n],\tau_{\mathrm{fb}})$, which is available from the onboard navigation unit.
The problem is formulated as
\begin{subequations}
\label{eq:OriginalOptimizationProblem}
\begin{align}
    \max_{\ell[n],\,\mathcal{B}[n]} \quad
    & \blue{\mathbb{E}\Big\{\bar R_{\mathrm{sum}}(\mathcal{B}[n],\ell[n];n)\,\Big|\,\widehat{\mathcal{I}}[n],v[n],\tau_{\mathrm{fb}}\Big\}}
    \label{eq:joint_problem_obj_journal}\\
    \text{s.t.}\quad
    & \ell[n]\in\{0,1,2,3\},
    \label{eq:joint_problem_c1_journal}\\
    & \mathcal{B}[n]\in\mathcal{P}_{\ell[n]}[n],
    \label{eq:joint_problem_c2_journal}\\
    & |\mathcal{B}[n]| = N_{\mathrm{RF}},
    \label{eq:joint_problem_c3_journal}\\
    & \tau_{\mathrm{tot}}(\ell[n])=\tau_{\mathrm{sw}}+\tau_{\mathrm{fb}}+\tau_{\mathrm{proc}}(\ell[n])+\tau_{\mathrm{p}},
    \label{eq:joint_problem_c4_journal}\\
    & \sum\nolimits_{k \in \mathcal{S}[n]}\left\|\mathbf{F}_{\mathrm{RF}}[n]\mathbf{f}_k[n] \right\|_2^2 \le P_t,
    \label{eq:joint_problem_c5_journal}
\end{align}
\end{subequations}
where $\mathcal{P}_{\ell[n]}[n]$ denotes the candidate pool associated with refinement level $\ell[n]$, and $\bar R_{\mathrm{sum}}(\mathcal{B}[n],\ell[n];n)$ is evaluated over the transmission channel at $t_{\mathrm{tx}}(\ell[n])$ with the payload fraction $\eta(\ell[n])$ of \eqref{eq:payload_fraction}.

Problem \eqref{eq:OriginalOptimizationProblem} highlights a key trade-off such that increasing the refinement level enlarges the candidate pool and may improve analog beam selection, but it also increases processing delay, shortens the payload, and aggravates the mobility-induced training-transmission mismatch.
Therefore, the optimal refinement level is not fixed a priori, and depends on both the beam-conflict structure revealed by the feedback and the severity of channel aging.

\vspace{-0.25cm}
\subsection{\blue{Feedback-Only Predicted Utility}}
\label{subsec:predicted_utility}

{\color{black}
The conditional expectation in \eqref{eq:joint_problem_obj_journal} is not available in closed form, so it is replaced by a predictor that consumes nothing beyond \eqref{eq:compact_feedback_packet}.
For a candidate subset $\mathcal{B}=\{b_1,\ldots,b_{N_{\mathrm{RF}}}\}$, the UAV assembles the predicted effective channel $\widehat{\mathbf{G}}(\mathcal{B},\ell)\in\mathbb{C}^{|\mathcal{S}|\times N_{\mathrm{RF}}}$ whose entries are read from the report and discounted by the predicted correlation, i.e.,
\begin{equation}
\big[\widehat{\mathbf{G}}(\mathcal{B},\ell)\big]_{k,i}
=
\begin{cases}
\rho(\ell,n)\,\widehat{c}_{k,j}[n], & b_i=\widehat{b}_{k,j}[n]\ \text{for some } j,\\[1mm]
0, & \text{otherwise},
\end{cases}
\label{eq:predicted_eff}
\end{equation}
where $\rho(\ell,n)$ follows from \eqref{eq:rho_model} with $\tau_{\mathrm{tot}}(\ell)$ of \eqref{eq:total_delay}.
The codewords that UE $k$ did not report are unknown rather than absent, and the innovation term of \eqref{eq:aged_channel_model} is unknown as well.
Both are collected into the per-UE residual power
\begin{equation}
\varepsilon_k^2(\mathcal{B},\ell)
=
\big(1-\rho^2(\ell,n)\big)\big|\widehat{c}_{k,1}[n]\big|^2
+
\zeta\,\big|\widehat{c}_{k,L_{\mathrm{r}}}[n]\big|^2 m_k(\mathcal{B}),
\label{eq:residual}
\end{equation}
in which $m_k(\mathcal{B})$ counts the codewords of $\mathcal{B}$ that UE $k$ did not report and $\zeta\in(0,1)$ is a leakage constant.
Applying RZF to $\widehat{\mathbf{G}}$ and normalizing the result to the transmit power gives the predicted stream matrix $\widehat{\mathbf{T}}=\widehat{\mathbf{G}}\widehat{\mathbf{F}}_{\mathrm{BB}}$, from which the predicted SINR is calculated as
\begin{equation}
\widehat{\gamma}_k(\mathcal{B},\ell)
=
\frac{\rho_{\mathrm{tx}}\big|[\widehat{\mathbf{T}}]_{k,k}\big|^2}
{1+\rho_{\mathrm{tx}}\sum\nolimits_{j\neq k}\big|[\widehat{\mathbf{T}}]_{k,j}\big|^2+\rho_{\mathrm{tx}}\,\varepsilon_k^2(\mathcal{B},\ell)},
\label{eq:predicted_sinr}
\end{equation}
and the predicted net utility of the pair $(\mathcal{B},\ell)$ is defined as
\begin{equation}
\widehat{U}(\mathcal{B},\ell)
=
\eta(\ell)\sum\nolimits_{k\in\mathcal{S}[n]}\log_2\big(1+\widehat{\gamma}_k(\mathcal{B},\ell)\big).
\label{eq:predicted_utility}
\end{equation}
Problem \eqref{eq:OriginalOptimizationProblem} is then replaced by its predictable counterpart
\begin{equation}
\big(\ell^{\star}[n],\mathcal{B}^{\star}[n]\big)
=
\arg\max_{\ell\in\{0,1,2,3\}}\;\max_{\mathcal{B}\in\mathcal{P}_{\ell}[n]}\;\widehat{U}(\mathcal{B},\ell),
\label{eq:predictable_problem}
\end{equation}
which is executable from the compact report alone.
Solving the inner maximization for a single level is affordable, whereas solving it for all four levels at every frame would defeat the purpose of adapting the depth, since the UAV would pay the cost of the deepest level in every frame.
Sections~\ref{subsec:conflict_metric_journal}--\ref{subsec:multilevel_refinement_decision_journal} therefore construct two scalar statistics of the report and use them to predict $\ell^{\star}[n]$ before any pool is built, and Section~\ref{sec:simulation_results} quantifies what this approximation costs against an oracle that evaluates all four levels.
}

%--------
\subsection{Conflict Metric from Beam-Training Feedback}
\label{subsec:conflict_metric_journal}

Given the analog beam subset $\mathcal{B}_{\mathrm{GS}}[n]$, the reliability of this baseline depends on the level of beam contention and ambiguity revealed by the reported beam indicators. 
To quantify this effect, we define a conflict score based on four lightweight components, including sector collisions, beam collisions, beam ambiguity, and beam completion requirement.

Let $n_q[n]$ denote the number of scheduled UEs whose strongest reported beam belongs to sector $q\in\{1,\ldots,N_s\}$, let $m_b[n]$ denote the number of them that report beam $b\in\{1,\ldots,M\}$ as their strongest beam, and let $\widetilde{\mathcal{B}}_{\mathrm{GS}}[n]$ denote the set of distinct strongest reported beams prior to beam completion.
The sector-collision count, the beam-collision count, the number of ambiguous UEs, and the number of beams required to complete the greedy set are respectively given by
\begin{align}
C_{\mathrm{sec}}[n] &= \sum\nolimits_{q=1}^{N_s}{n_q[n]\big(n_q[n]-1\big)}/{2},
\label{eq:sector_collision_journal_new}\\
C_{\mathrm{beam}}[n] &= \sum\nolimits_{b=1}^{M}{m_b[n]\big(m_b[n]-1\big)}/{2},
\label{eq:beam_collision_journal_new}\\
C_{\mathrm{amb}}[n] &= \sum\nolimits_{k \in {\cal S}[n]}
\mathbb{1}\!\left(\widehat{\Delta}_k[n] < \Delta_{\mathrm{th}}\right),
\label{eq:ambiguity_count_journal_new}\\
C_{\mathrm{fill}}[n] &= \max\!\big(0,\,N_{\mathrm{RF}}-\big|\widetilde{\mathcal{B}}_{\mathrm{GS}}[n]\big|\big),
\label{eq:fill_count_journal_new}
\end{align}
where $\Delta_{\mathrm{th}}$ is a predefined ambiguity threshold and $\widehat{\Delta}_k[n]$ is the quantized beam-gap indicator read from \eqref{eq:compact_feedback_packet}.
Beam ambiguity is therefore characterized by the gap between the strongest and second-strongest reported beams, and beam completion by the number of RF chains that no UE has claimed.

{\color{black}
The four counts live on different scales and grow differently with the system size: the two collision counts scale as $|\mathcal{S}[n]|^2$, the ambiguity count as $|\mathcal{S}[n]|$, and the completion count as $N_{\mathrm{RF}}$.
A fixed weighted sum of the raw counts would therefore not be invariant to $K$, $M$, $N_s$, or $N_{\mathrm{RF}}$, and the same thresholds could not be reused when the load changes.
Moreover, a beam collision is by construction also a sector collision, so the two counts overlap and the same event would be charged twice.
Each component is accordingly normalized by its own maximum, the beam collisions are removed from the sector count, and the weights are normalized to sum to one, i.e.,
\begin{equation}
% \resizebox{\linewidth}{!}{$
S_{\mathrm{conf}}[n]
\!=\!
\frac{\omega_1 \bar{C}_{\mathrm{sec}}[n]
\!+\omega_2 \bar{C}_{\mathrm{beam}}[n]
\!+\omega_3 \bar{C}_{\mathrm{amb}}[n]
\!+\omega_4 \bar{C}_{\mathrm{fill}}[n]}
{\omega_1+\omega_2+\omega_3+\omega_4},
% $}
\label{eq:conflict_score}
\end{equation}
with $S=|\mathcal{S}[n]|$,
$\bar{C}_{\mathrm{sec}}[n]=\frac{C_{\mathrm{sec}}[n]-C_{\mathrm{beam}}[n]}{S(S-1)/2}$,
$\bar{C}_{\mathrm{beam}}[n]=\frac{C_{\mathrm{beam}}[n]}{S(S-1)/2}$,
$\bar{C}_{\mathrm{amb}}[n]=\frac{C_{\mathrm{amb}}[n]}{S}$,
and
$\bar{C}_{\mathrm{fill}}[n]=\frac{C_{\mathrm{fill}}[n]}{N_{\mathrm{RF}}}$,
% \begin{align}
% \bar{C}_{\mathrm{sec}}[n]&=\frac{C_{\mathrm{sec}}[n]-C_{\mathrm{beam}}[n]}{S(S-1)/2}, &
% \bar{C}_{\mathrm{beam}}[n]&=\frac{C_{\mathrm{beam}}[n]}{S(S-1)/2}, \nonumber\\
% \bar{C}_{\mathrm{amb}}[n]&=\frac{C_{\mathrm{amb}}[n]}{S}, &
% \bar{C}_{\mathrm{fill}}[n]&=\frac{C_{\mathrm{fill}}[n]}{N_{\mathrm{RF}}},
% \label{eq:conflict_components}
% \end{align}
where $\omega_1,\omega_2,\omega_3,$ and $\omega_4$ are nonnegative weighting coefficients.
For $S<2$, the normalized sector- and beam-collision terms are defined as zero.
The score therefore takes values in $[0,1]$ for every system size, and the thresholds calibrated in Section~\ref{subsec:multilevel_refinement_decision_journal} transfer across $K$ without retuning, which is verified in Section~\ref{sec:simulation_results}.
}
A larger $S_{\mathrm{conf}}[n]$ indicates that the GS beam subset is more likely to be suboptimal, suggesting that deeper candidate refinement may be beneficial.

%----------
\subsection{Pre-Decision Aging Severity}
\label{subsec:aging_severity_journal}

In addition to beam conflict, the benefit of candidate refinement depends on mobility-induced channel aging. While a richer candidate search can improve beam selection accuracy on the training channel, it also incurs additional processing delay and may become less effective when UAV mobility is high or the feedback delay is large.
To capture this trade-off, we define an aging-severity metric as
\begin{equation}
A[n]
=
\alpha_v {v[n]}/{v_{\mathrm{ref}}}
+
\alpha_{\tau}{\tau_{\mathrm{fb}}}/{\tau_{\mathrm{ref}}},
\label{eq:aging_severity}
\end{equation}
where
% $v[n]$ denotes the UAV speed, $\tau_{\mathrm{fb}}$ is the feedback delay,
$v_{\mathrm{ref}}$ and $\tau_{\mathrm{ref}}$ are reference speed and delay, respectively, which are normalization constants, while coefficients $\alpha_v,\alpha_{\tau}\ge 0$ are weighting factors.
A larger value of $A[n]$ indicates more severe training-transmission mismatch and thus discourages aggressive refinement.

{\color{black}
Both arguments of \eqref{eq:aging_severity} are known to the UAV without any signaling, since $v[n]$ is read from the onboard navigation unit and $\tau_{\mathrm{fb}}$ is a configured system parameter, which is what makes $A[n]$ usable at decision time.
The metric is a linear surrogate for the expected loss suffered by a beam that was selected on the training snapshot and is used at $t_{\mathrm{tx}}$, i.e.,
\begin{equation}
\Delta G(v,\tau)
=
\mathbb{E}\Big\{10\log_{10}\frac{g^{\mathrm{tr}}_{k,\widehat{b}_{k,1}}\big(t_{\mathrm{tx}}\big)}{\max_m g^{\mathrm{tr}}_{k,m}\big(t_{\mathrm{tx}}\big)}\Big\},
\label{eq:stale_gain_loss}
\end{equation}
which is the gain that the reported beam still delivers relative to the best beam at the transmission instant.
Two distinct mechanisms drive $\Delta G$, namely the decorrelation of the diffuse component, which is governed by $f_{\mathrm{D}}^{\max}\tau_{\mathrm{tot}}$ and therefore by $v$, and the angular drift produced by the attitude dynamics \eqref{eq:attitude}, whose peak rate also grows with $v$.
Both mechanisms are monotone in $v$ and in $\tau_{\mathrm{fb}}$, and Section~\ref{sec:simulation_results} verifies that $\Delta G$ measured on the geometry-based model is monotone in $A[n]$, so a scalar summary is sufficient for ordering the refinement levels.
The linear form is retained because $(\alpha_v,\alpha_\tau)$ and the thresholds are calibrated jointly, which absorbs the shape of $\Delta G$ into the threshold values.
}

%-------
\subsection{Multi-Level Refinement Decision}
\label{subsec:multilevel_refinement_decision_journal}

To enable a more flexible refinement mechanism, the proposed design allows multiple refinement levels. 
% Let $\ell[n]\in\mathcal{L}=\{0,1,2,3\}$ denote the refinement level selected at transmission interval $n$. Here, $\ell[n]=0$ corresponds to the GS baseline without refinement, $\ell[n]=1$ corresponds to sector-top-$2$ refinement, $\ell[n]=2$ corresponds to global top-$2$ refinement, and $\ell[n]=3$ corresponds to global top-$3$ refinement. A larger refinement level therefore corresponds to a richer candidate search and potentially higher beam-selection accuracy, at the expense of additional processing cost and delay.
%
The refinement level is determined by the joint effect of conflict severity and aging severity. 
In particular, higher conflict scores favor deeper refinement, whereas more severe aging favors shallower refinement. 
Let $(\tau_1,\tau_2,\tau_3)$ and $(A_1,A_2)$ denote increasing conflict thresholds and aging thresholds, respectively, satisfying $\tau_1 <\tau_2 <\tau_3$ and $A_1<A_2$. 
The multi-level decision is given by
\begin{equation}
\ell[n]=\Psi\!\left(S_{\mathrm{conf}}[n],A[n]\right),
\label{eq:level_mapping_general}
\end{equation}
where $\Psi(\cdot)$ denotes a rule-based mapping such that
\begin{equation}
\ell[n]
=
\begin{cases}
0, & S_{\mathrm{conf}}[n] < \tau_1,\\[1mm]
1, & \tau_1 \le S_{\mathrm{conf}}[n] < \tau_2,\; A[n] < A_2,\\[1mm]
0, & \tau_1 \le S_{\mathrm{conf}}[n] < \tau_2,\; A[n] \ge A_2,\\[1mm]
2, & \tau_2 \le S_{\mathrm{conf}}[n] < \tau_3,\; A[n] < A_1,\\[1mm]
1, & \tau_2 \le S_{\mathrm{conf}}[n] < \tau_3,\; A_1 \le A[n] < A_2,\\[1mm]
3, & S_{\mathrm{conf}}[n] \ge \tau_3,\; A[n] < A_1,\\[1mm]
2, & S_{\mathrm{conf}}[n] \ge \tau_3,\; A_1 \le A[n] < A_2,\\[1mm]
1, & S_{\mathrm{conf}}[n] \ge \tau_3,\; A[n] \ge A_2.
\end{cases}
\label{eq:level_mapping_piecewise}
\end{equation}
This formulation reflects the core design principle of the proposed framework such that deeper refinement is selected only when beam conflict is sufficiently severe and mobility-induced aging remains moderate.

{\color{black}
The map in \eqref{eq:level_mapping_piecewise} is a calibrated two-statistic approximation motivated by the predictable selection problem in \eqref{eq:predictable_problem}. Writing
$\widehat{U}^{\star}(\ell)
=
\max_{\mathcal{B}\in\mathcal{P}_{\ell}[n]}
\widehat{U}(\mathcal{B},\ell)$,
the predictable level selector would evaluate
$\ell^{\star}[n]
=
\arg\max_{\ell}\widehat{U}^{\star}(\ell)$.
Performing this maximization over all four levels at every frame, however, would require constructing and evaluating every candidate pool and would therefore remove the computational benefit of adaptive refinement.

The online mapping instead uses two statistics that summarize complementary effects available before candidate construction. The conflict score $S_{\mathrm{conf}}[n]$ measures the observable degree of MU beam contention, ambiguity, and beam-completion demand in the compact report. A smaller conflict score provides less evidence that an expanded candidate search is needed, whereas a larger score indicates greater potential benefit from refinement. In contrast, the aging metric $A[n]$ summarizes the expected loss associated with mobility, feedback latency, and the reduced payload available after a deeper search. These two trends motivate the monotone structure of \eqref{eq:level_mapping_piecewise}: the selected refinement depth is non-decreasing with conflict severity and non-increasing with aging severity. No claim is made that this two-statistic mapping exactly solves \eqref{eq:predictable_problem} for every individual frame.}

\blue{The thresholds are obtained offline by a design-stage calibration. A training set of channel trajectories, disjoint in seed from the evaluation set and covering $v\in[0,80]$~m/s and $\tau_{\mathrm{fb}}\in[0,5]$~ms, is generated, all four levels are executed on every frame, and the pair $(\boldsymbol{\tau},\mathbf{A})$ that maximizes the realized proportional-fair network utility $\sum_{k\in\mathcal{S}[n]}\log\bar{R}_k[n]$ is retained by grid search. 
While \eqref{eq:predictable_problem} defines the per-frame beam-selection utility for a fixed scheduled set, the threshold calibration uses the proportional-fair network utility to account for the long-term scheduling objective. 
The proportional-fair utility is adopted because it is consistent with the scheduler and rewards protection of poorly served UEs, for which refinement is particularly useful. Only the offline calibration uses the realized rates. During online operation, the policy observes only the compact report and the mobility state. Section~\ref{sec:simulation_results} reports the fitted thresholds, the resulting agreement with the per-frame optimal level, and a sensitivity study of all five thresholds and of the conflict weights.}

Given the selected level, the UAV builds the corresponding candidate pool and picks the subset that maximizes the predicted utility, which is detailed in Section~\ref{subsec:candidate_evaluation_selection}.
The proposed design therefore does not rely on a fixed trigger, and a single-level triggered strategy is the special case of \eqref{eq:level_mapping_general} in which only one refinement level is available once refinement is activated.

%------
\section{Single-Level Conflict-Triggered Refinement as a Special Case}
\label{sec:tg3tr_revisited}

Before delving into the proposed multi-level policy, we first describe a single-level conflict-triggered refinement strategy that motivates the overall framework.
The key idea is to retain the GS solution as the default low-complexity mode and activate a deeper candidate search only when the reported beam indicators suggest that the current realization is conflict-prone.
This leads to a single-level triggered global top-3 refinement strategy, referred to as TG3-TR, in which the UAV computes a conflict score from compact beam-training feedback and invokes a fixed global top-3 candidate search when the score exceeds a predefined threshold ${\tau_{\mathrm{trig}}}$.
The corresponding decision and beam subset are respectively given by
\begin{equation}
u_{\mathrm{TG3}}[n]
=
\begin{cases}
0, & S_{\mathrm{conf}}[n] < {\tau_{\mathrm{trig}}},\\[1mm]
1, & S_{\mathrm{conf}}[n] \ge {\tau_{\mathrm{trig}}},
\end{cases}
\label{eq:tg3tr_trigger}
\end{equation}
\begin{equation}
\mathcal{B}_{\mathrm{TG3}}[n]
\!=\!
\begin{cases}
\mathcal{B}_{\mathrm{GS}}[n], & u_{\mathrm{TG3}}[n]=0,\\[1mm]
\displaystyle
\arg\max_{\mathcal{B}[n]\in\mathcal{P}_{3}[n]}
\blue{\widehat{U}(\mathcal{B}[n],3)}, & u_{\mathrm{TG3}}[n]=1,
\end{cases}
\label{eq:tg3tr_beam_selection}
\end{equation}
where the candidate ranking uses the same feedback-only utility as the proposed policy, so that TG3-TR and MLR-TG differ only in how the refinement depth is chosen.
\blue{Setting ${\tau_{\mathrm{trig}}}=\tau_2$ and disabling levels $1$ and $2$ in \eqref{eq:level_mapping_piecewise} recovers \eqref{eq:tg3tr_trigger}, so TG3-TR is the two-level restriction of the proposed map rather than a separate design.}

\begin{remark}[Baseline and Motivation]
    Notably, TG3-TR preserves the low-complexity GS solution under typical channel conditions and activates a richer candidate search only when the conflict severity exceeds the trigger threshold\footnote{The trigger threshold is set offline by the same calibration procedure as the multi-level thresholds. A smaller threshold activates refinement more frequently and brings TG3-TR closer to always-on refinement, whereas a larger threshold reduces the activation rate and makes the method closer to the GS baseline. For a fair comparison, it is aligned with the intermediate conflict threshold used in MLR-TG.}, i.e., $S_{\mathrm{conf}}[n] \ge {\tau_{\mathrm{trig}}}$.
    Despite its simplicity, TG3-TR is inherently limited by its fixed refinement depth.
    \blue{A global top-3 search may be unnecessarily costly when the conflict is mild, and under high mobility or a large feedback delay the same depth is overly aggressive, because it lengthens the decision latency, shortens the payload through \eqref{eq:payload_fraction}, and increases the training-transmission mismatch}.
    Furthermore, when beam contention is severe under moderate aging, a binary on/off refinement strategy cannot exploit the structure of the reported beam candidates.
    These limitations motivate the more general framework in which the UAV determines not only whether refinement should be activated but also the appropriate refinement depth, thereby conceptualizing the proposed MLR-TG in Section~\ref{sec:proposed_method}.
\end{remark}

\blue{TG3-TR is retained as a benchmark in Section~\ref{sec:simulation_results} and is executed with Algorithm~\ref{alg:mlrtg} restricted to $\ell[n]\in\{0,3\}$, so no separate listing is required.}

\vspace{-0.25cm}
%==============
\section{Proposed Mobility- and Feedback-Aware Multi-Level Refinement Policy}
\label{sec:proposed_method}

% This section presents the proposed mobility- and feedback-aware MLR-TG policy. 
% The policy operates in two stages. 
% First, the UAV constructs the GS beam subset from compact beam-training feedback and evaluates the conflict severity and aging severity for the current transmission interval. 
% Then, based on the selected refinement level, the UAV either retains the GS solution or activates a level-dependent candidate search and selects the final analog beam subset that maximizes the achievable sum-SE. 
% This design preserves the low complexity of GS beamforming in typical or aging-dominant scenarios, while enabling progressively stronger refinement when beam conflict is severe and the expected channel mismatch remains moderate.
This section presents the proposed MLR-TG policy, which operates in two stages: the UAV first evaluates the conflict and aging severities to select a refinement level, and then either retains the GS solution or performs a level-dependent candidate search. This preserves the low complexity of GS beamforming in typical or aging-dominant scenarios, while enabling stronger refinement when beam conflict is severe and the expected mismatch remains moderate.

\vspace{-0.25cm}
\subsection{GS Initialization and Level Selection}
\label{subsec:gs_init_level_selection}

% At frame $n$, the UAV extracts the strongest reported beam of each UE from the compact feedback set $\widehat{\mathcal{I}}[n]$ and constructs the GS beam subset $\mathcal{B}_{\mathrm{GS}}[n]$ as described in Section~\ref{subsec:gs_and_multilevel_refinement}. This subset serves as the default low-complexity analog beamforming solution.
At frame $n$, the UAV constructs the GS beam subset $\mathcal{B}_{\mathrm{GS}}[n]$ from $\widehat{\mathcal{I}}[n]$ as described in Section~\ref{subsec:gs_and_multilevel_refinement}, which serves as the default low-complexity solution.

%Subsequently, the UAV computes the conflict score $S_{\mathrm{conf}}[n]$ from \eqref{eq:conflict_score} and the aging-severity metric $A[n]$ from \eqref{eq:aging_severity}. The refinement level is then determined via the rule-based mapping in \eqref{eq:level_mapping_general}. This mapping reflects the intuition that when the reported beam indicators exhibit limited contention or the expected channel aging is severe, the UAV avoids costly candidate expansion and selects a shallow refinement level. In contrast, when the beam conflict is pronounced and aging remains moderate\footnote{\color{black}From which equation that determines they are "moderate"? it is good to be clear}, the policy selects a deeper refinement level to improve analog beam selection. Consequently, the proposed method does not treat refinement as a binary on/off decision. Alternatively, it adapts the refinement depth across transmission intervals, which distinguishes MLR-TG from single-level triggered designs.

Subsequently, the UAV computes the conflict score $S_{\mathrm{conf}}[n]$ from \eqref{eq:conflict_score} and the aging-severity metric $A[n]$ from \eqref{eq:aging_severity}, and determines the refinement level through the mapping in \eqref{eq:level_mapping_piecewise}.
Limited contention, i.e. $S_{\mathrm{conf}}[n]<\tau_1$, or severe aging, i.e. $A[n]\geq A_2$, lead to a shallow level, whereas the deepest level is reached only when $S_{\mathrm{conf}}[n]\geq\tau_3$ and $A[n]<A_1$.
Refinement is therefore not a binary on/off decision, which is what distinguishes MLR-TG from single-level triggered designs.

\vspace{-0.25cm}
\subsection{Level-Dependent Candidate Pool Construction}
\label{subsec:level_dependent_pool}

For different refinement levels $\ell[n]$, the UAV constructs corresponding candidate pools. 
It is worth mentioning that the case $\ell[n] = 0$ is described in Section~\ref{subsec:gs_and_multilevel_refinement}, while $\ell[n] \in \{ 1,2,3 \}$ is detailed below.

% Let $\widehat{q}_k[n]$, $\widehat{\mathcal{T}}_k[n]$, and $\widehat{b}_k[n]$ denote the reported dominant sector, candidate beam list, and strongest beam of user $k$, respectively.

Level $\ell[n]=1$ performs a sector top-$2$ refinement, in which the candidate set of UE $k$ is restricted to the two strongest beams within its reported dominant sector, whereas levels $\ell[n]=2$ and $\ell[n]=3$ explore the two and three strongest reported beams across all sectors. The three candidate sets are respectively expressed as
\begin{align}
\mathcal{U}_k^{(1)}[n] &\subseteq \mathcal{C}_{\widehat{q}_k[n]},
\quad \big|\mathcal{U}_k^{(1)}[n]\big|\le 2,
\label{eq:level1_user_options}\\
\mathcal{U}_k^{(2)}[n] &= \{\text{two strongest beams in } \widehat{\mathcal{T}}_k[n]\},
\label{eq:level2_user_options}\\
\mathcal{U}_k^{(3)}[n] &= \{\text{three strongest beams in } \widehat{\mathcal{T}}_k[n]\},
\label{eq:level3_user_options}
\end{align}
so that level $1$ provides a lightweight refinement inside the most relevant angular region, level $2$ enables inter-sector exploration at the same cardinality, and level $3$ is the richest mode considered.

For a given refinement level $\ell[n]\in\{1,2,3\}$, a raw beam combination is formed by selecting one beam from each UE-specific set $\mathcal{U}_k^{(\ell[n])}[n]$. 
Specifically,
\begin{equation}
\widetilde{\mathcal{B}}_{\ell[n]}^{(i)}[n]
=
\left\{
u_1^{(i)}[n],\ldots,u_K^{(i)}[n]
\right\},
\\
u_k^{(i)}[n]\in\mathcal{U}_k^{(\ell[n])}[n],
\label{eq:raw_combination}
\end{equation}
denotes the $i$-th raw combination. 
Duplicate beams within each combination are removed. 
If the number of distinct beams is smaller than $N_{\mathrm{RF}}$, the remaining RF chains are filled using unused beams with the largest aggregate training gain such that
\begin{equation}
\bar{g}_m[n]
=
\sum\nolimits_{k \in {\cal K}} g_{k,m}^{\mathrm{tr}}[n], \,\,
\forall m \in \{1,\ldots,M \}.
\label{eq:aggregate_training_gain}
\end{equation}
After duplicate removal and beam completion, a valid beam subset of cardinality $N_{\mathrm{RF}}$ is obtained,
which forms the candidate pool for $\ell [n]$ as
% $\mathcal{P}_{\ell[n]}[n]
% =
% \left\{
% \widetilde{\mathcal{B}}_{\ell[n]}^{(1)}[n],\ldots,
% \widetilde{\mathcal{B}}_{\ell[n]}^{(N_{\mathrm{cand}}(\ell[n],n))}[n]
% \right\}$.
% Let $\mathcal{P}_{\ell[n]}[n]$ denote the resulting set of unique valid candidates after removing repeated subsets. Then
\begin{equation}
\mathcal{P}_{\ell[n]}[n]
=
\left\{
\widetilde{\mathcal{B}}_{\ell[n]}^{(1)}[n],\ldots,
\widetilde{\mathcal{B}}_{\ell[n]}^{(N_{\mathrm{cand}}(\ell[n],n))}[n]
\right\}.
\label{eq:level_candidate_pool}
\end{equation}
Here, $N_{\mathrm{cand}}(\ell[n],n)$ denotes the number of retained candidates.
Before duplicate removal, the raw search size is at most $2^{|\mathcal{S}[n]|}$ for $\ell[n] \in \{ 1,2 \}$, and at most $3^{|\mathcal{S}[n]|}$ for $\ell[n]=3$.
{\color{black}
An exhaustive product would therefore become intractable well before the array does, so the enumeration is bounded rather than complete.
Indexing the options of UE $k$ by $i_k\in\{1,\ldots,|\mathcal{U}_k^{(\ell)}[n]|\}$ in decreasing reported gain, combinations are generated in increasing order of the total deviation $\sum_{k}(i_k-1)$ from the per-UE strongest option and the enumeration stops at a hard cap $N_{\max}$, i.e.,
\begin{equation}
\mathcal{P}_{\ell}[n]
\subseteq
\Big\{\widetilde{\mathcal{B}}:\textstyle\sum_{k\in\mathcal{S}[n]}(i_k-1)\le D_{\ell}[n]\Big\},
\quad
\big|\mathcal{P}_{\ell}[n]\big|\le N_{\max},
\label{eq:pool_cap}
\end{equation}
where $D_{\ell}[n]$ is the largest deviation radius that fits within the cap.
Combinations that deviate from the greedy assignment in few positions are thus explored first, which matches the structure of the problem, since a conflict typically involves a small number of contending UEs.
The complete product is recovered whenever it fits, which is the case for levels $1$ and $2$ in the configuration of Section~\ref{sec:simulation_results}.}
Therefore, the refinement level directly controls the candidate pool dimension and, via \eqref{eq:total_delay}, the processing delay.

% [Search Region of Each Refinement Level]
\begin{remark}
The candidate pools corresponding to different refinement levels differ in \blue{where they search as well as in how widely they search}.
Specifically, sector top-$2$ refinement performs a local search within the reported dominant sector, global top-$2$ refinement expands the search to the two strongest beams across sectors, and global top-$3$ refinement further enlarges the candidate space.
Therefore, higher refinement levels provide richer search spaces at the cost of increased processing complexity and delay.
\end{remark}

\begin{proposition}[\blue{Monotonicity and Exact Reduction}]\label{Proposition:1}
\blue{Let $\Psi(S_{\mathrm{conf}},A)$ denote the map in \eqref{eq:level_mapping_piecewise}. Then}
\begin{enumerate}[(i)]
\item \blue{for every fixed $S_{\mathrm{conf}}$, $\Psi(S_{\mathrm{conf}},A)$ is non-increasing in $A$, and for every fixed $A$, $\Psi(S_{\mathrm{conf}},A)$ is non-decreasing in $S_{\mathrm{conf}}$;}
\item \blue{MLR-TG reduces to the GS baseline, i.e., $\Psi(S_{\mathrm{conf}},A)=0$, if and only if $S_{\mathrm{conf}}<\tau_1$, or $\tau_1\le S_{\mathrm{conf}}<\tau_2$ and $A\ge A_2$;}
\item \blue{MLR-TG coincides with always-on global top-$3$ refinement, i.e., $\Psi(S_{\mathrm{conf}},A)=3$, if and only if $S_{\mathrm{conf}}\ge\tau_3$ and $A<A_1$.}
\end{enumerate}
\begin{proof}
\blue{Fix $S_{\mathrm{conf}}$ and read \eqref{eq:level_mapping_piecewise} column by column. For $S_{\mathrm{conf}}<\tau_1$, the level is $0$ for all $A$. For $\tau_1\le S_{\mathrm{conf}}<\tau_2$, the level takes the values $1$ and $0$ on $A<A_2$ and $A\ge A_2$, respectively. For $\tau_2\le S_{\mathrm{conf}}<\tau_3$, it takes $2$ and $1$ on $A<A_1$ and $A\ge A_1$. For $S_{\mathrm{conf}}\ge\tau_3$, it takes $3$, $2$, and $1$ on $A<A_1$, $A_1\le A<A_2$, and $A\ge A_2$. Each column is a non-increasing step function of $A$, which gives the first half of (\textit{i}). Fixing $A$ and reading row by row over the four conflict intervals yields the sequences $(0,1,1,1)$ for $A\ge A_2$, $(0,1,1,2)$ for $A_1\le A<A_2$, and $(0,1,2,3)$ for $A<A_1$, each non-decreasing in $S_{\mathrm{conf}}$, which gives the second half of (\textit{i}).} Statements (\textit{ii}) and (\textit{iii}) collect the cells of \eqref{eq:level_mapping_piecewise} whose value is $0$ and $3$, respectively, and are exhaustive by construction.
\end{proof}
\end{proposition}

\begin{remark}[\blue{Interpretation Under Severe Aging}]
\blue{A large aging severity does not force the greedy mode in general. For $S_{\mathrm{conf}}\ge\tau_3$ and $A\ge A_2$, the map returns $\ell=1$, and for $\tau_2\le S_{\mathrm{conf}}<\tau_3$ and $A_1\le A<A_2$, it also returns $\ell=1$, so a severe conflict always retains at least the cheapest refinement level.
The exact property enforced by the design is the monotonicity of Proposition~\ref{Proposition:1}(i), namely that the selected depth never increases with aging severity and never decreases with conflict severity.}
\end{remark}

\vspace{-0.25cm}
%-----
\subsection{Candidate Evaluation and Final Beam Selection}
\label{subsec:candidate_evaluation_selection}

For each candidate beam subset $\mathcal{B}[n]\in\mathcal{P}_{\ell[n]}[n]$, the UAV constructs the corresponding analog precoder as
\begin{align}\label{eq:frf_candidate}
\mathbf{F}_{\mathrm{RF}}(\mathcal{B}[n];n)
&=
\big[
\mathbf{c}_{b_1[n]},
\mathbf{c}_{b_2[n]},
\dots,
\mathbf{c}_{b_{N_{\mathrm{RF}}}[n]}
\big],
\\
\mathcal{B}[n]&=\{b_1[n],\dots,b_{N_{\mathrm{RF}}}[n]\}.
\end{align}
{\color{black}
Candidates are then ranked, and only the winner is realized.
Ranking every candidate with $R_{\mathrm{sum}}$ would require the transmission-time channel, or an effective channel measured separately for each candidate, neither of which is available before the analog stage is committed.
The ranking therefore uses the feedback-only utility \eqref{eq:predicted_utility}, and is organized in two stages so that the exact predictor is applied only where it matters.
The first stage scores all $|\mathcal{P}_{\ell[n]}[n]|$ candidates with the closed-form surrogate obtained by replacing $\widehat{\mathbf{T}}$ in \eqref{eq:predicted_sinr} with the reported gain of the codeword assigned to each UE, weighting the gains of the remaining codewords of the subset by a residual factor $\kappa_{\mathrm{res}}\in(0,1)$ that stands for the interference the digital stage will not remove, and charging the full gain of a codeword that two UEs would have to share, which costs ${\cal O}(|\mathcal{S}[n]|L_{\mathrm{r}})$ operations per candidate.
The second stage retains the $N_{\mathrm{top}}(\ell[n])$ best distinct subsets, assembles $\widehat{\mathbf{G}}$ from \eqref{eq:predicted_eff} for each of them, and evaluates \eqref{eq:predicted_utility} exactly, so that
\begin{equation}
\mathcal{B}^{\star}[n]
=
\begin{cases}
\mathcal{B}_{\mathrm{GS}}[n], & \ell[n]=0,\\[1mm]
\displaystyle
\arg\max_{\mathcal{B}[n]\in\mathcal{P}^{\mathrm{top}}_{\ell[n]}[n]}
\widehat{U}\big(\mathcal{B}[n],\ell[n]\big), & \ell[n]\in\{1,2,3\},
\end{cases}
\label{eq:final_beam_selection_repeated}
\end{equation}
where $\mathcal{P}^{\mathrm{top}}_{\ell}[n]\subseteq\mathcal{P}_{\ell}[n]$ collects the promoted subsets and $N_{\mathrm{top}}(\ell)$ grows with the level.
Once $\mathcal{B}^{\star}[n]$ is fixed, the pilot stage of \eqref{eq:heff_estimate} is executed once and the digital precoder is computed by
\begin{align}
&\mathbf{F}_{\mathrm{BB}}(\mathcal{B}^{\star}[n];n)
=
\widehat{\mathbf{H}}_{\mathrm{eff}}^H(\mathcal{B}^{\star}[n];n)\nonumber\\
&\qquad\times
\Big(
\widehat{\mathbf{H}}_{\mathrm{eff}}(\mathcal{B}^{\star}[n];n)
\widehat{\mathbf{H}}_{\mathrm{eff}}^H(\mathcal{B}^{\star}[n];n)
+\mu\mathbf{I}_{|\mathcal{S}[n]|}
\Big)^{-1},
\label{eq:fbb_candidate}
\end{align}
followed by normalization to satisfy the transmit power constraint in \eqref{eq:power_constraint}.
% Exactly one RZF inversion and one pilot stage are executed per frame regardless of the selected level, which is what keeps the information model and the overhead accounting consistent.
Exactly one pilot stage and one transmission RZF inversion based on the pilot-estimated effective CSI are executed per frame regardless of the selected level. Any additional regularized inversions used for feedback-only candidate prediction are confined to the promoted subsets and are explicitly accounted for in $C_{\mathrm{ex}}$.
}
The resulting analog and digital precoders of the MLR-TG are expressed, respectively, as
\begin{equation}
\mathbf{F}_{\mathrm{RF}}^{\star}[n]
=
\mathbf{F}_{\mathrm{RF}}(\mathcal{B}^{\star}[n];n),
\label{eq:final_frf_star}
\end{equation}
\begin{equation}
\mathbf{F}_{\mathrm{BB}}^{\star}[n]
=
\mathbf{F}_{\mathrm{BB}}(\mathcal{B}^{\star}[n];n).
\label{eq:final_fbb_star}
\end{equation}
Notably, $\mathbf{F}_{\mathrm{BB}}^{\star}[n]$ in \eqref{eq:final_fbb_star} is also normalized to guarantee \eqref{eq:power_constraint}.
This procedure indicates that level selection and candidate evaluation are tightly coupled. When the aging severity is high, the policy favors shallow refinement or reverts to the GS baseline. In contrast, when the conflict score is high and mismatch remains moderate, the policy expands the candidate space and adopts deeper refinement.

% This procedure implies that level selection and candidate evaluation are tightly coupled. If the aging severity is high, the policy tends to restrict the search to a shallow refinement level or even keep the GS baseline. If the conflict score is large and the expected mismatch remains moderate, the policy expands the candidate space and uses a stronger refinement mode.

% Algorithm~\ref{alg:mlrtg} summarizes the proposed MLR-TG procedure. The policy starts from compact beam-training feedback, computes the conflict score and aging severity, selects an appropriate refinement level, and then performs level-dependent candidate construction and RZF-based candidate evaluation. In this way, the proposed method adapts the beam-selection effort to the current operating condition rather than invoking the same refinement intensity for every interval.

\begin{algorithm}[t]
\caption{Proposed MLR-TG: Mobility- and Feedback-Aware Multi-Level Refinement}
\label{alg:mlrtg}
\KwIn{\blue{Compact feedback $\widehat{\mathcal{I}}[n]$, UAV speed $v[n]$, feedback delay $\tau_{\mathrm{fb}}$, codebook $\mathcal{C}$, sector partition $\{\mathcal{C}_q\}_{q=1}^{N_s}$, and calibrated thresholds $\tau_1,\tau_2,\tau_3,A_1,A_2$.}}
\KwOut{Final analog precoder $\mathbf{F}_{\mathrm{RF}}^{\star}[n]$ and digital precoder $\mathbf{F}_{\mathrm{BB}}^{\star}[n]$.}
Construct GS beam subset $\mathcal{B}_{\mathrm{GS}}[n]$ based on \eqref{eq:GS_baseline}\;
Compute conflict score $S_{\mathrm{conf}}[n]$ according to \eqref{eq:conflict_score}\;
Compute aging-severity metric $A[n]$ according to \eqref{eq:aging_severity}\;
Determine refinement level $\ell[n]$ using \eqref{eq:level_mapping_piecewise}\;

\eIf{$\ell[n]=0$} {
    Set $\mathcal{B}^{\star}[n]\leftarrow \mathcal{B}_{\mathrm{GS}}[n]$\; 
}
{
    \blue{Build the bounded-deviation candidate pool $\mathcal{P}_{\ell[n]}[n]$ using \eqref{eq:pool_cap}}\;
    \blue{Score every candidate with the closed-form surrogate and promote the $N_{\mathrm{top}}(\ell[n])$ best distinct subsets to $\mathcal{P}^{\mathrm{top}}_{\ell[n]}[n]$}\;
    \For{\upshape \blue{each candidate $\mathcal{B}[n]\in\mathcal{P}^{\mathrm{top}}_{\ell[n]}[n]$}} {
        \blue{Assemble the predicted effective channel $\widehat{\mathbf{G}}(\mathcal{B}[n],\ell[n])$ using \eqref{eq:predicted_eff} and the residual power using \eqref{eq:residual}}\;
        \blue{Evaluate the predicted net utility $\widehat{U}(\mathcal{B}[n],\ell[n])$ using \eqref{eq:predicted_utility}}\;
    }
    Set $\mathcal{B}^{\star}[n]\leftarrow \blue{\arg\max_{\mathcal{B}[n]\in\mathcal{P}^{\mathrm{top}}_{\ell[n]}[n]} \widehat{U}(\mathcal{B}[n],\ell[n])}$\;
}
Return final analog precoder $\mathbf{F}_{\mathrm{RF}}^{\star}[n]$ according to \eqref{eq:final_frf_star}\;
\blue{Transmit $N_{\mathrm{p}}$ pilots through $\mathbf{F}_{\mathrm{RF}}^{\star}[n]$ and acquire $\widehat{\mathbf{H}}_{\mathrm{eff}}[n]$ using \eqref{eq:heff_estimate}}\;
Compute digital precoder $\mathbf{F}_{\mathrm{BB}}^{\star}[n]$ using \eqref{eq:final_fbb_star}, normalize itself to satisfy \eqref{eq:power_constraint}, and return it.
\end{algorithm}

%Algorithm~\ref{alg:mlrtg} summarizes the proposed MLR-TG procedure. Starting from compact beam-training feedback, the UAV computes the conflict score and aging severity, selects a refinement level, and performs level-dependent candidate construction followed by RZF-based evaluation. This adaptive strategy adjusts the refinement depth to current operating conditions, rather than applying a fixed refinement intensity across intervals.

\subsection{\color{black}Overall Algorithm and Complexity Analysis}
\blue{Algorithm~\ref{alg:mlrtg} summarizes the proposed MLR-TG procedure. Starting from compact beam-training feedback, the UAV computes the conflict score and aging severity, selects a refinement level, and performs level-dependent candidate construction followed by feedback-only predicted-utility evaluation. The transmission RZF precoder is formed only after the final analog beam subset has been selected. This adaptive strategy adjusts the refinement depth to current operating conditions rather than applying a fixed refinement intensity across frames.}

% [Refinement Depth Versus Latency]
\begin{remark}
A deeper candidate search does not necessarily improve end-to-end performance under UAV mobility. Although higher refinement levels may improve beam-selection accuracy based on the reported training information, they also incur longer processing delays and thus larger training--transmission mismatch. The proposed MLR-TG policy explicitly exploits this trade-off by reducing the refinement depth when aging severity becomes significant.
\end{remark}

{\color{black}
%The complexity of MLR-TG is derived below from Algorithm~\ref{alg:mlrtg} with the matrix dimensions written out, counting a complex multiply-accumulate as eight real FLOPs and setting $S=|\mathcal{S}[n]|$. Every mode pays the same per-frame cost and only the refinement term depends on the level, which corrects an accounting error of the conference version in which the common term was charged to the greedy mode alone.
The complexity of MLR-TG is derived below from Algorithm~\ref{alg:mlrtg} with the matrix dimensions written out, counting a complex multiply-accumulate as eight real floating-point operations (FLOPs) and setting $S=|\mathcal{S}[n]|$. Every refinement mode is charged the same common per-frame cost, while only the level-dependent candidate-evaluation term varies with the selected refinement depth.
The common term collects the UE-side codebook correlation $\mathcal{O}(8KMN_t)$, the greedy construction and beam completion $\mathcal{O}(2KS+N_{\mathrm{RF}}M)$, the conflict and aging metrics $\mathcal{O}(6K+2N_s)$, the effective-channel formation $\mathcal{O}(8SN_{\mathrm{RF}}N_{\mathrm{p}})$, and the single post-selection transmission-RZF stage $\mathcal{O}(16S^2N_{\mathrm{RF}}+\tfrac{8}{3}S^3)$, i.e.,
\begin{equation}
C_{\mathrm{com}}
\!=\!
{\cal O}\big(8KMN_t+N_{\mathrm{RF}}M+8SN_{\mathrm{RF}}N_{\mathrm{p}}+16S^2N_{\mathrm{RF}}+\tfrac{8}{3}S^3\big).
\label{eq:complexity_common}
\end{equation}
The refinement term of level $\ell$ comprises the coarse scoring of $\bar N_\ell$ candidates, at $C_{\mathrm{cs}}={\cal O}(S(3L_{\mathrm{r}}+4)+S\log_2 S+2N_{\mathrm{RF}}+\tfrac{1}{4}N_{\mathrm{RF}}M)$ FLOPs each, and the exact evaluation of $N_{\mathrm{top}}(\ell)$ promoted subsets, at $C_{\mathrm{ex}}={\cal O}(SL_{\mathrm{r}}N_{\mathrm{RF}}+24S^2N_{\mathrm{RF}}+\tfrac{20}{3}S^3)$ FLOPs each, the latter covering the assembly of $\widehat{\mathbf{G}}$, its Gram matrix, the regularized inverse, and the predicted stream matrix.
The resulting number of FLOPs for MLR-TG is obtained as a weighted combination of the four modes, i.e.,
\begin{equation}
{\bar C}_{\mathrm{MLR}}
=
C_{\mathrm{com}}
+
\sum\nolimits_{\ell=1}^{3}\varrho_\ell\big(\bar N_\ell C_{\mathrm{cs}} + N_{\mathrm{top}}(\ell)C_{\mathrm{ex}}\big),
\label{eq:complexity_mlr}
\end{equation}
where $0 \leq \varrho_\ell \leq 1$ denotes the long-term fraction of frames in which refinement level $\ell$ is selected, satisfying $\sum_{\ell=0}^{3} \varrho_\ell = 1$, and $\bar N_1$, $\bar N_2$, and $\bar N_3$ are the average pool cardinalities written as $\bar N_{S2}$, $\bar N_{G2}$, and $\bar N_{G3}$ elsewhere.
Both $\varrho_\ell$ and $\bar N_\ell$ are measured in Section~\ref{sec:simulation_results}, and setting $\varrho_3=1$ or $\varrho_0=1$ recovers always-on global top-$3$ refinement (AG3) or pure GS, so the three codebook-based schemes are compared on one common accounting.}
\blue{The decision latency of \eqref{eq:total_delay} follows from the same count, i.e., $\tau_{\mathrm{proc}}(\ell)=\tau_{\mathrm{pipe}}+C_{\mathrm{ref}}(\ell)/\Theta$, where $\Theta$ is the sustained throughput of the onboard processor, stated with $\tau_{\mathrm{pipe}}$ in Table~\ref{tab:simulation_parameters}, and the sensitivity of the conclusions to $\Theta$ is examined through a latency-scaling sweep.}

% This structure highlights that MLR-TG does not always incur the cost of the deepest refinement level. Instead, it adaptively allocates refinement effort based on conflict and aging conditions, achieving a more favorable performance-complexity trade-off than always-on global top-3 refinement.

% In terms of feedback overhead, MLR-TG employs the same compact beam-training packet as defined in \eqref{eq:compact_feedback_packet}.
% The refinement level only affects the extent of candidate evaluation and does not require additional high-dimensional CSI feedback. This makes the proposed design suitable for mobility-sensitive UAV systems operating under limited and delayed feedback.

MLR-TG therefore does not always incur the cost of the deepest level. It also employs the same compact packet \eqref{eq:compact_feedback_packet} at every level, since the refinement level affects only the extent of candidate evaluation and requires no additional high-dimensional CSI feedback.

% \textcolor{red}{Check until this...}

\vspace{-0.25cm}
\section{Simulation Results and Discussions}
\label{sec:simulation_results}

\subsection{Simulation Parameters and Benchmark Schemes}

{\color{black}
In this section, we present SLSs to evaluate the proposed framework under the causal timeline of Section~\ref{subsec:frame_timeline}.
Independent topologies are drawn, each comprising a UAV heading, a UE deployment, scatterer positions, shadowing draws, and Clarke phases, and each is simulated over $20$ consecutive beam-management frames so that the channel evolves continuously and the Kalman benchmark can maintain a state.
The main operating point uses $250$ topologies, i.e. $5000$ frames, whereas the parameter sweeps use $100$ and the design studies $60$.
All schemes see the same topologies, trajectory, and pilot-noise draws, and differ only in the instant at which they transmit.
Every parameter is listed in Table~\ref{tab:simulation_parameters}, the information available to each scheme in Table~\ref{tab:assumptions}, and the aging constant $\beta$ is fitted to the geometry-based model rather than assumed.
}

\begin{table}[t]
\caption{\blue{Simulation parameters.}}
\label{tab:simulation_parameters}
\centering
\color{black}
\renewcommand{\arraystretch}{1.05}
\resizebox{\linewidth}{!}{
\begin{tabular}{|p{3cm}|>{\centering\arraybackslash}p{2.75cm}||p{3cm}|>{\centering\arraybackslash}p{2cm}|}
\hline
\textbf{Parameter} & \textbf{Value} & \textbf{Parameter} & \textbf{Value} \\
\hline\hline
Carrier, bandwidth & $28$~GHz, $100$~MHz & UAV altitude $H_{\mathrm{U}}$ & $100$~m \\\hline
Antennas $N_t$ & $128$ & Cell radius $R_{\mathrm{c}}$, $r_{\min}$ & $200$, $20$~m \\\hline
RF chains $N_{\mathrm{RF}}$ & $8$ & UAV speed $v$ & $40$~m/s \\\hline
UEs $K$ & $8$ & Rician factor $\kappa_{\mathrm{R}}$ & $10$~dB \\\hline
Paths $L$ & $3$ & Path-loss exp. LoS/NLoS & $2.0$ / $3.0$ \\\hline
Codebook $M$, sectors $N_s$ & $128$, $8$ & NLoS excess loss & $20$~dB \\\hline
Report size $L_{\mathrm{r}}$ & $5$ & Shadowing LoS/NLoS & $4$ / $6$~dB \\\hline
$(q_{\mathrm{A}},q_{\mathrm{P}},q_{\mathrm{eff}})$ & $(5,5,10)$ bits & Hotspots $(\varrho_{\mathrm{h}},N_{\mathrm{h}},r_{\mathrm{h}})$ & $(0.5,2,15\,\mathrm{m})$ \\\hline
Symbol $T_{\mathrm{s}}$ (120~kHz subcarrier spacing) & $8.93~\mu$s & Attitude $(\omega_0,\omega_1;f_{\mathrm{a}})$ & $(0.15,1.5)$ rad/s; $2$ Hz \\\hline
Frame $T_{\mathrm{f}}$, delay $\tau_{\mathrm{fb}}$ & $10$, $1$~ms & Clarke osc. $N_{\mathrm{o}}$, scatterer ring & $8$, $15$~m \\\hline
Pilots $N_{\mathrm{p}}$, boost $\rho_{\mathrm{p}}$ & $8$, $10$ & Transmit SNR $\rho_{\mathrm{tx}}$ & $5$~dB \\\hline
Processor $\Theta$, $\tau_{\mathrm{pipe}}$ & $2$~GFLOP/s, $60~\mu$s & SINR threshold $\gamma_{\mathrm{th}}$ & $1$~dB \\\hline
Pool cap $N_{\max}$; $N_{\mathrm{top}}(1\!:\!3)$ & $1024$; $(8,16,32)$ & Gap threshold $\Delta_{\mathrm{th}}$; $\epsilon_{\mathrm{pf}}$ & $3$~dB; $0.05$ \\\hline
Leakage $\zeta$ & $0.5$ & Frames per topology & $20$ \\\hline
Conflict weights $(\omega_1,\omega_2,\omega_3,\omega_4)$ & $(1,\,1.5,\,0.75,\,0.5)$ & Residual interference $\kappa_{\mathrm{res}}$ & $0.12$ \\\hline
Aging references $(v_{\mathrm{ref}},\tau_{\mathrm{ref}})$ & $(50~\mathrm{m/s},\,2~\mathrm{ms})$ & Aging weights $(\alpha_v,\alpha_\tau)$ & $(1,\,1)$ \\\hline
Calibrated $(\tau_1,\tau_2,\tau_3)$ & $(0.04,\,0.16,\,0.23)$ & Calibrated $(A_1,A_2)$; fitted $\beta$ & $(0.90,\,2.30)$; $7.12\!\times\!10^{-3}$ \\\hline
\end{tabular}}
%\vspace{-0.35cm}
\end{table}

\begin{table*}[t]
\caption{\blue{Information, feedback, timing, and complexity assumptions of each scheme. $C_{\mathrm{com}}$, $C_{\mathrm{cs}}$, and $C_{\mathrm{ex}}$ are defined in \eqref{eq:complexity_common} and Section~\ref{sec:proposed_method}, and $S=|\mathcal{S}[n]|$.}}
\label{tab:assumptions}
\centering
\small
\color{black}
\renewcommand{\arraystretch}{1.05}
\resizebox{\linewidth}{!}{
\begin{tabular}{|p{1.5cm}|p{7.0cm}|p{2.1cm}|p{4.3cm}|c|c|}
\hline
\textbf{Scheme} & \textbf{Knowledge used for the analog stage} & \textbf{Uplink report} & \textbf{Computational complexity (FLOPs)} & \textbf{Avg. FLOPs} & \textbf{$\tau_{\mathrm{proc}}$ ($\mu$s)} \\
\hline\hline
FD-RZF & full instantaneous CSI at its own transmit instant (genie) & $N_t$ complex coeff. & $\mathcal{O}(8N_tS^2+\tfrac{8}{3}S^3+8N_tS)$ & $7.5\!\times\!10^{4}$ & $98$ \\\hline
OMP-RZF & full CSI acquired at the training instant, then aged & $N_t$ complex coeff. & $\mathcal{O}(8N_{\mathrm{RF}}M_{\mathrm{OMP}}N_tS)+C_{\mathrm{com}}$ & $8.9\!\times\!10^{6}$ & $4523$ \\\hline
GS & strongest reported index only & $1$ index & $C_{\mathrm{com}}$ & $1.5\!\times\!10^{4}$ & $68$ \\\hline
KF-BT & strongest reported index plus a per-UE extended Kalman filter (EKF) state & $1$ index & $C_{\mathrm{com}}+\mathcal{O}(40K)$ & $1.5\!\times\!10^{4}$ & $68$ \\\hline
TG3-TR & compact report \eqref{eq:compact_feedback_packet} & $L_{\mathrm{r}}$ index-coeff. pairs & $C_{\mathrm{com}}+\varrho_{\mathrm{TG3}}(\bar N_{G3}C_{\mathrm{cs}}+N_{\mathrm{top}}(3)C_{\mathrm{ex}})$ & $5.9\!\times\!10^{4}$ & $68$/$553$ \\\hline
AG3 & compact report \eqref{eq:compact_feedback_packet} & $L_{\mathrm{r}}$ index-coeff. pairs & $C_{\mathrm{com}}+\bar N_{G3}C_{\mathrm{cs}}+N_{\mathrm{top}}(3)C_{\mathrm{ex}}$ & $9.9\!\times\!10^{5}$ & $553$ \\\hline
\textbf{MLR-TG} & compact report \eqref{eq:compact_feedback_packet} & $L_{\mathrm{r}}$ index-coeff. pairs & $C_{\mathrm{com}}+\sum_{\ell=1}^{3}\varrho_\ell(\bar N_\ell C_{\mathrm{cs}}+N_{\mathrm{top}}(\ell)C_{\mathrm{ex}})$, i.e. \eqref{eq:complexity_mlr} & $2.3\!\times\!10^{5}$ & $68$--$553$ \\\hline
Oracle-L / Oracle-C & noncausal CSI at each level's own transmit instant, and under the common level-$0$ delay & genie & --- & --- & per level / $68$ \\\hline
\end{tabular}}
\end{table*}

%\blue{All schemes acquire the effective channel through the same pilot stage \eqref{eq:heff_estimate} and build RZF from it, so Table~\ref{tab:assumptions} differs only in the analog stage. The two oracles are not implementable and serve as bounds.}
\blue{The implementable codebook-based schemes, namely GS, KF-BT, TG3-TR, AG3, and MLR-TG, use the same post-selection effective-channel acquisition stage in \eqref{eq:heff_estimate} and construct the transmission RZF precoder from the resulting estimate. FD-RZF and OMP-RZF retain the full-CSI assumptions stated explicitly in Table~\ref{tab:assumptions}, while Oracle-L and Oracle-C are noncausal bounds.}

For benchmark comparisons, the proposed MLR-TG, implemented via Algorithm~\ref{alg:mlrtg}, is evaluated against \blue{eight} schemes, including
fully digital regularized zero-forcing (\textbf{FD-RZF}),
orthogonal matching pursuit based HBF with RZF digital precoding (\textbf{OMP-RZF}),
pure GS beamforming,
\blue{Kalman-filter beam tracking (\textbf{KF-BT}),}
triggered global top-$3$ refinement (\textbf{TG3-TR}),
always-on global top-$3$ refinement (\textbf{AG3})\blue{, and two oracles}.
FD-RZF \cite{nguyen2019multi} serves as a fully digital upper bound without analog beamforming constraints.
OMP-RZF \cite{noh2016zero} represents a conventional sparse HBF approach, where the analog precoder is obtained via orthogonal matching pursuit (OMP) over the same $M$-codeword dictionary and the digital precoder via RZF. \blue{Accordingly, $M_{\mathrm{OMP}}=M$ denotes the OMP dictionary size, and the OMP analog stage performs $N_{\mathrm{RF}}$ atom-selection iterations.}
Pure GS follows \cite[Alg.~1]{wu2023simultaneous}, which is a low-complexity baseline that directly selects beams from the strongest reported UE beams with beam completion when necessary.
\blue{KF-BT is the mobility-aware benchmark motivated by the beam-tracking literature \cite{hyun2021adaptive,8811740,11422871}. Each UE runs a two-state extended Kalman filter on the direction cosine of its dominant path, driven by the codeword index with measurement variance $(2/M)^2/12$, and the UAV serves the codeword closest to the state predicted at $t_{\mathrm{tx}}$, so KF-BT anticipates the drift that the codebook-based schemes absorb at the same reporting cost as GS.}
TG3-TR activates a fixed global top-3 refinement only for conflict-prone realizations, \blue{using the trigger \eqref{eq:tg3tr_trigger} with ${\tau_{\mathrm{trig}}}=\tau_2$}. \blue{In the complexity expression of Table~\ref{tab:assumptions}, ${\varrho_{\mathrm{TG3}}}\in[0,1]$ denotes the long-term fraction of frames in which this trigger is active; it equals $0.045$ at the main operating point.}
In contrast, AG3 applies global top-3 refinement in every \blue{frame}, i.e., $\ell =3$ is always selected in Algorithm~\ref{alg:mlrtg}.
{\color{black}
The two oracles isolate the two sources of loss separately, namely level selection and candidate restriction.
\textbf{Oracle-L} evaluates the winner of every level on the transmission-time channel of that level and keeps the best, which upper bounds what any level-selection rule can achieve under the same physical delays.
\textbf{Oracle-C} evaluates the $32$ most promising level-$3$ candidates on the transmission-time channel under the common level-$0$ delay and keeps the best, which upper bounds what any candidate-selection rule can achieve once the delay is removed.
Because Oracle-C searches a subset of the pool, it is itself a lower bound on the unrestricted oracle.
}

% The proposed MLR-TG scheme is compared with five benchmark methods, namely fully digital regularized zero-forcing (FD-RZF), orthogonal matching pursuit based hybrid beamforming with RZF digital precoding (OMP-RZF), GS beamforming (GS), triggered global top-$3$ refinement (TG3-TR), and always-on global top-$3$ refinement (AG3). 
% Here, FD-RZF serves as a fully digital upper reference without analog beamforming constraints. OMP-RZF represents a conventional sparse hybrid beamforming benchmark in which the analog precoder is obtained through orthogonal matching pursuit and the digital precoder is constructed by RZF. GS is the default low-complexity baseline that directly forms the analog beam subset from the strongest reported user beams with beam completion when necessary. TG3-TR is a single-level conflict-triggered refinement strategy that activates a fixed global top-$3$ candidate search only for conflict-prone realizations. AG3 performs the same global top-$3$ refinement for every transmission interval and therefore serves as an upper benchmark for refinement-based codebook beam selection. 
% In contrast, MLR-TG adaptively selects the refinement depth from the current conflict severity and aging severity, and can operate as the greedy baseline, a sector-top-$2$ refinement, a global top-$2$ refinement, or a global top-$3$ refinement depending on the operating condition.

\vspace{-0.25cm}
\subsection{Performance Evaluation}

{\color{black}
Four metrics are reported, namely the net sum-SE of \eqref{eq:net_se} averaged over frames, the raw sum-SE $R_{\mathrm{sum}}$ that separates the airtime cost from the beam-selection quality, the $5$th percentile of the per-UE net rate, and the system OP, i.e. the empirical fraction of frames in which at least one scheduled UE falls below the target SINR,
\begin{equation}
P_{\mathrm{out}}
=
\frac{1}{N_{\mathrm{f}}}\sum\nolimits_{n=1}^{N_{\mathrm{f}}}
\mathbb{1}\Big(\min\nolimits_{k\in\mathcal{S}[n]}\gamma_k[n] < \gamma_{\mathrm{th}}\Big),
\label{eq:outage_def}
\end{equation}
where $N_{\mathrm{f}}$ is the number of simulated frames and the threshold is converted to the linear scale before use, i.e. $\gamma_{\mathrm{th}}=10^{1/10}=1.259$, which corresponds to a per-UE SE threshold of $\log_2(1+\gamma_{\mathrm{th}})=1.176$~bps/Hz.}
\blue{Fairness is measured by Jain's index $J=(\sum_{k\in\mathcal{K}}\bar R_k)^2/(K\sum_{k\in\mathcal{K}}\bar R_k^{\,2})$ of the long-term average net rates $\bar R_k$ accumulated over a topology, which counts the zero rates of the UEs that were not scheduled and therefore measures fairness across time rather than within a frame.}

\blue{Before turning to the schemes, the model abstractions are validated.
Measured on the geometry-based model at $v=40$~m/s, the phase-invariant correlation of the reported beamspace signature decays from $1$ to $0.929$ after $1$~ms and to $0.893$ after $5$~ms, whereas the correlation of the raw complex channel vector has already fallen to $0.376$ after $1$~ms, which is why the analog stage tolerates a delay that the digital stage does not.
A least-squares fit of \eqref{eq:rho_model} to the beamspace correlation gives $\beta=7.12\times10^{-3}$ with a normalized root-mean-square error of $0.027$, and this fitted value is used throughout instead of an assumed one.
Sweeping $v\in[0,80]$~m/s and $\tau_{\mathrm{fb}}\in[0,5]$~ms, the mean gain loss of the reported beam at the transmission instant, defined in \eqref{eq:stale_gain_loss}, spans $0$ to $-3.36$~dB with a rank correlation of $0.875$ against $A[n]$, so a scalar orders the operating points closely enough for the levels to be ranked.
The offline calibration of Section~\ref{subsec:multilevel_refinement_decision_journal} returns $(\tau_1,\tau_2,\tau_3)=(0.04,0.16,0.23)$ and $(A_1,A_2)=(0.90,2.30)$, and the fitted rule reproduces the per-frame optimal level in $42.7\%$ of the calibration frames while losing only $3.59\%$ of the oracle utility, because adjacent levels are frequently close and picking the second best is inexpensive.
}

\begin{figure}[t]
    \centering
    \includegraphics[width=\linewidth]{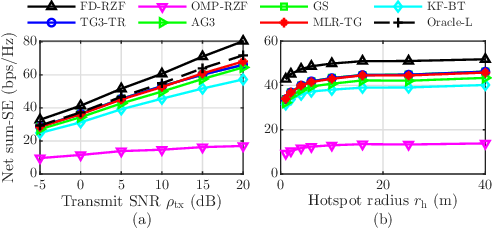}
    \caption{\blue{Average net sum-SE versus (a) the transmit SNR and (b) the hotspot radius $r_{\mathrm{h}}$.}}
    \label{fig:se_snr}
\end{figure}

{\color{black}
Figs.~\ref{fig:se_snr}(a) and \ref{fig:se_snr}(b) show the average net sum-SE versus the transmit SNR and versus the hotspot radius, respectively.
As shown in Fig.~\ref{fig:se_snr}(a), the net sum-SE of every scheme grows with the SNR, and FD-RZF remains the highest throughout because it is granted full instantaneous CSI and $N_t$ RF chains.
Among the codebook-based schemes, MLR-TG and GS are within $1\%$ of each other below $10$~dB, while MLR-TG becomes the strongest for $\rho\ge15$~dB and exceeds GS by $3.3\%$ at $20$~dB, because resolving a beam conflict is an interference effect that only becomes visible once the system leaves the noise-limited regime.
In particular, AG3 is $6.1\%$ below GS at $5$~dB although its raw sum-SE is $0.3\%$ above it, which is the airtime it spends on a global top-$3$ search in every frame.
OMP-RZF achieves the lowest net sum-SE of all schemes, at $13.9$~bps/Hz at $5$~dB, for two reasons that Table~\ref{tab:assumptions} makes explicit: its dictionary search costs $8.9\times10^{6}$ FLOPs, i.e. $45\%$ of the frame on the assumed processor, and its greedy atom selection matches the fully digital precoder as a whole rather than allocating one codeword per UE, so weakly served UEs are left without a dedicated beam.
KF-BT reaches $86\%$ of the GS net sum-SE at $5$~dB.
Fig.~\ref{fig:se_snr}(b) shows that the net sum-SE of all schemes grows as the hotspot radius increases from $1$ to $40$~m, over which the mean intra-hotspot separation of the direction cosines grows from $0.17$ to $6.65$ beamwidths, the mean conflict score falls from $0.112$ to $0.074$, and the mean level selected by MLR-TG falls from $1.01$ to $0.84$.
The reliability gain moves in the opposite direction, since MLR-TG reduces the OP with respect to GS by only $4.1\%$ at $r_{\mathrm{h}}=1$~m but by $21.8\%$ at $r_{\mathrm{h}}=40$~m, because UEs packed inside a single beamwidth cannot be separated by any codeword assignment.
}

\begin{figure}[t]
    \centering
    \includegraphics[width=\linewidth]{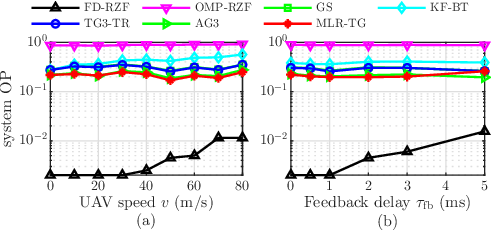}
    \caption{\blue{System outage probability versus (a) the UAV speed and (b) the feedback delay.}}
    \label{fig:outage_velocity}
    % \vspace{-0.3cm}
\end{figure}

{\color{black}
Figs.~\ref{fig:outage_velocity}(a) and \ref{fig:outage_velocity}(b) show the system OP versus the UAV speed and the feedback delay, respectively. As shown in Fig.~\ref{fig:outage_velocity}(a), MLR-TG achieves the lowest OP among the implementable limited-feedback schemes over the entire speed range, reducing the OP relative to GS by $21.1\%$ at $v=0$ and $29.9\%$ at $v=80$~m/s. In comparison, the OPs of GS and TG3-TR remain nearly unchanged with speed, which is consistent with the relatively slow decay of the beamspace correlation reported above. The OP of KF-BT increases from $0.272$ to $0.566$, while its net sum-SE decreases from $45.4$ to $34.5$~bps/Hz. This degradation occurs because its constant-velocity state model captures the smooth translational motion but not the yaw and pitch corrections in \eqref{eq:attitude}, whose rate increases with airspeed. Hence, angle prediction alone becomes less effective under the considered rotary-wing attitude dynamics. Fig.~\ref{fig:outage_velocity}(b) shows that MLR-TG maintains the lowest OP up to $\tau_{\mathrm{fb}}=3$~ms. At $\tau_{\mathrm{fb}}=5$~ms, the aging severity becomes sufficiently large that MLR-TG selects the greedy mode in $98.5\%$ of frames. Its OP consequently increases to $0.259$, while its net sum-SE of $21.23$~bps/Hz becomes nearly identical to that of GS. At the same operating point, AG3 achieves a $24.8\%$ lower OP, but at the cost of a $13.7\%$ reduction in net sum-SE. This result illustrates the intended behavior of MLR-TG: because the policy is calibrated using the proportional-fair utility of the net rates, it avoids deeper refinement when the additional reliability gain no longer compensates for the associated airtime loss.
}

\begin{figure}[t]
    \centering
    \includegraphics[width=\linewidth]{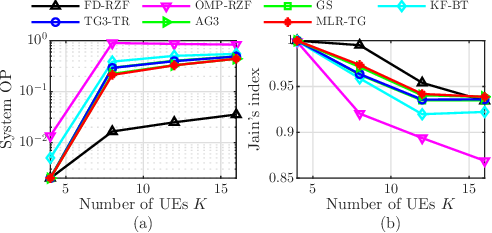}
    \caption{\blue{Impact of the number of UEs on (a) the system outage probability and (b) Jain's fairness index, with $N_{\mathrm{RF}}=8$ and proportional-fair scheduling for $K>N_{\mathrm{RF}}$.}}
    \label{fig:user_scaling_outage_fairness}
%    \vspace{-0.3cm}
\end{figure}

{\color{black}
Figs.~\ref{fig:user_scaling_outage_fairness}(a) and \ref{fig:user_scaling_outage_fairness}(b) show the effect of the UE load on the system OP and on Jain's index, respectively, with $N_{\mathrm{RF}}=8$ fixed so that $K=12$ and $K=16$ exercise the scheduler of Section~\ref{subsec:gs_and_multilevel_refinement}.
The thresholds calibrated at $K=8$ are reused unchanged at every load, which is possible because the conflict score of \eqref{eq:conflict_score} is normalized component by component.
As shown in Fig.~\ref{fig:user_scaling_outage_fairness}(a), the OP of every scheme grows with $K$ because more UEs compete for the same angular resources, and MLR-TG reduces it with respect to GS by $27.7\%$, $18.8\%$, and $10.3\%$ at $K=8$, $12$, and $16$. The shrinking margin follows from the load, since at $K=16$ the scheduler already discards half the UEs and the remaining conflicts are the ones no codeword assignment can resolve.
Fig.~\ref{fig:user_scaling_outage_fairness}(b) shows that MLR-TG attains the highest Jain index among the codebook-based schemes at every load, at $0.9735$ against $0.9629$ for GS at $K=8$, and the load-normalized net rate per scheduled UE, $5.63$ against $5.69$~bps/Hz, confirms that the fairness gain is not obtained by starving the strong UEs.
}

\begin{figure}[t]
    \centering
    \includegraphics[width=\linewidth]{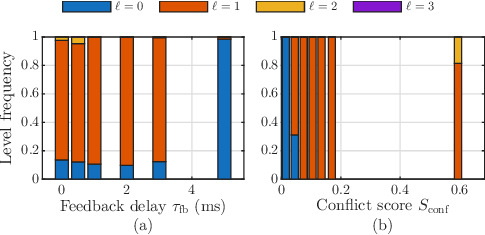}
    \caption{\blue{Frequency with which each refinement level is selected, versus (a) the feedback delay and (b) the conflict score.}}
    \label{fig:levels}
%    \vspace{-0.3cm}
\end{figure}

{\color{black}
Fig.~\ref{fig:levels} reports the level-selection statistics that make the adaptive behavior visible.
The level distribution is nearly invariant to the UAV speed, with the greedy mode selected in $9.3\%$ to $14.2\%$ of frames over $v\in[0,80]$~m/s, consistent with the slow decay of the beamspace correlation, whereas Fig.~\ref{fig:levels}(a) shows a sharp transition in the feedback delay, since refinement is active in roughly $88\%$ of frames up to $\tau_{\mathrm{fb}}=3$~ms and collapses to $1.5\%$ at $\tau_{\mathrm{fb}}=5$~ms, where the aging severity crosses $A_2$.
Fig.~\ref{fig:levels}(b) shows the mechanism that drives the choice, since the selected level increases with the conflict score and the two deepest levels appear only in the most contended frames.
Averaged over the main operating point the policy selects $\ell=0$, $1$, $2$, and $3$ in $11.4\%$, $88.3\%$, $0.3\%$, and less than $0.1\%$ of frames, and the rarity of $\ell=3$ is itself a result rather than an omission, since AG3 shows what selecting it in every frame costs.
}

\begin{figure}[t]
    \centering
    \includegraphics[width=\linewidth]{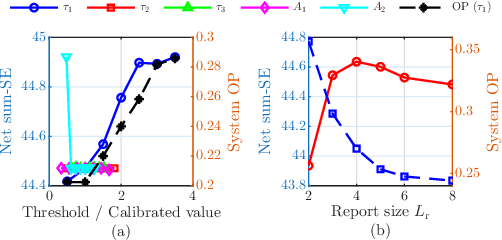}
    \caption{\blue{Design study of the proposed policy on net sum-SE and system OP with the impacts of (a) thresholds and (b) report sizes.}}
    \label{fig:design}
   \vspace{-0.25cm}
\end{figure}

{\color{black}
Fig.~\ref{fig:design}(a) sweeps each of the five thresholds around its calibrated value with common random numbers across settings, so that the differences are paired.
Raising $\tau_1$ from $0.02$ to $0.14$, i.e. refining in fewer frames, raises the net sum-SE by only $1.1\%$ but raises the OP by $41.1\%$ and lowers the $5$th-percentile rate by $43.6\%$, which is the trade-off the calibration resolves and the reason the fitted $\tau_1$ sits at the reliability-favouring end of the range.
Sweeping $A_2$ exposes the same switch from the aging side, since at $A_2=1.1$ the aging severity of the operating point exceeds the threshold, refinement is disabled altogether, and the $5$th-percentile rate falls from $2.10$ to $1.19$~bps/Hz, whereas any $A_2\ge1.4$ leaves the policy unchanged.
The three remaining thresholds, $\tau_2$, $\tau_3$, and $A_1$, leave all metrics unchanged to within $0.1\%$ over their sweeps, because the conflict score exceeds $\tau_2$ in fewer than $5\%$ of frames at this operating point, so they only become active in the most contended frames.
Varying the beam-collision weight $\omega_2$ over $[0.5,3]$ changes the net sum-SE by $0.26\%$ and the OP by $1.6\%$.
A two-dimensional sweep over the $(\tau_2,A_1)$ plane confirms the same conclusion, since the net sum-SE varies by $0.45\%$ and the OP by $10.9\%$ of their means across it.
The policy is therefore insensitive to three of its five thresholds and to the conflict weights in this regime, and the two that matter, $\tau_1$ and $A_2$, act as on/off switches that trade sum rate against reliability rather than degrading both.
Fig.~\ref{fig:design}(b) sweeps the report size.
The net sum-SE saturates at $L_{\mathrm{r}}=4$ to within $0.1\%$, whereas the OP keeps improving, from $0.357$ at $L_{\mathrm{r}}=2$ to $0.244$ at $L_{\mathrm{r}}=8$.
We adopt $L_{\mathrm{r}}=5$ because it captures $92\%$ of the achievable outage reduction at $1344$ bits per frame, which is $77\%$ of the $1752$ bits required by $L_{\mathrm{r}}=8$.
Sweeping the coefficient resolution separately, the net sum-SE saturates at $q_{\mathrm{A}}=q_{\mathrm{P}}=5$~bits and changes by less than $0.05\%$ beyond it, while dropping to $2$~bits costs $2.7\%$.
}

\begin{table}[t]
\color{black}
\caption{\blue{Beam-training and feedback overhead per beam-management frame, for $K=8$, $N_{\mathrm{RF}}=8$, $M=128$, $N_t=128$, and $L_{\mathrm{r}}=5$.}}
\label{tab:overhead}
\centering
\renewcommand{\arraystretch}{1.05}
\resizebox{\linewidth}{!}{
\begin{tabular}{|l|c|>{\centering\arraybackslash}p{2.5cm}|>{\centering\arraybackslash}p{1.8cm}|c|}
\hline
\textbf{Quantity} & \textbf{GS, KF-BT} & \textbf{TG3-TR, AG3, MLR-TG} & \textbf{Full beam report} & \textbf{Full CSI} \\
\hline\hline
Beam-sweep symbols & $128$ & $128$ & $128$ & $128$ \\\hline
Pilot symbols for $\mathbf{H}_{\mathrm{eff}}$ & $8$ & $8$ & $8$ & $8$ \\\hline
Beam report per UE (bits) & $7$ & $88$ & $768$ & $1280$ \\\hline
Beam report, $K$ UEs (bits) & $56$ & $704$ & $6144$ & $10240$ \\\hline
Effective-CSI report (bits) & $640$ & $640$ & $640$ & $0$ \\\hline
Total per frame (bits) & $696$ & $1344$ & $6784$ & $10240$ \\\hline
Reduction w.r.t. full CSI & $93.2\%$ & $86.9\%$ & $33.8\%$ & --- \\\hline
\end{tabular}}
%\vspace{-0.3cm}
\end{table}

{\color{black}
Table~\ref{tab:overhead} quantifies the claim that the report is compact.
The beam sweep and the pilot stage are identical for every scheme, whereas the reporting stage is not.
The proposed packet costs $88$~bits per UE and $1344$~bits per frame, which is $86.9\%$ below full-CSI reporting, $80.2\%$ below the full beam-quality report with $q_g=6$ bits per codeword, and $93.1\%$ below full CSI when only the beam-management report is compared. Relative to the strongest-index-only report used by GS and KF-BT, the compact packet adds $648$~bits per frame. This reporting cost is fixed and does not depend on the refinement level selected in a given frame.
}

\begin{table}[t]
\color{black}
\centering
\caption{\blue{Average performance at the main operating point ($\rho=5$~dB, $v=40$~m/s, $\tau_{\mathrm{fb}}=1$~ms, $K=N_{\mathrm{RF}}=8$, $5000$ frames).}}
\label{tab:performance_comparison}
\renewcommand{\arraystretch}{1.05}
\resizebox{\linewidth}{!}{
\begin{tabular}{|p{1.7cm}|c|c|c|c|c|c|}
\hline
\multirow{2}{*}{\textbf{Scheme}} & \textbf{Net sum-SE} & \textbf{Raw sum-SE} & \multirow{2}{*}{\textbf{System OP}} & \textbf{Jain's} & \textbf{5th-pct rate} & \textbf{Feedback} \\
& \textbf{(bps/Hz)} & \textbf{(bps/Hz)} & & \textbf{index} & \textbf{(bps/Hz)} & \textbf{(bits/frame)} \\
\hline\hline
\textbf{FD-RZF}   & 51.373 & 67.449 & 0.0064 & 0.9965 & 5.182 & 10240 \\ \hline
\textbf{OMP-RZF}  & 13.541 & 42.438 & 0.8824 & 0.8867 & 0.014 & 10240 \\ \hline
\textbf{Pure GS}  & 45.319 & 59.267 & 0.2814 & 0.9665 & 1.214 & 696 \\ \hline
\textbf{KF-BT}    & 39.015 & 51.023 & 0.4034 & 0.9612 & 0.698 & 696 \\ \hline
\textbf{TG3-TR}   & 45.259 & 59.331 & 0.2814 & 0.9667 & 1.207 & 1344 \\ \hline
\textbf{AG3}      & 42.554 & 59.426 & 0.2200 & 0.9731 & 1.632 & 1344 \\ \hline
\textbf{Proposed MLR-TG} & \textbf{44.886} & \textbf{59.528} & \textbf{0.2064} & \textbf{0.9742} & \textbf{1.868} & 1344 \\ \hline\hline
\textit{Oracle-L} & 46.465 & 61.293 & 0.2084 & 0.9768 & 2.341 & genie \\ \hline
\textit{Oracle-C} & 46.872 & 61.299 & 0.1170 & 0.9858 & 3.879 & genie \\ \hline
\end{tabular}}
%\vspace{-0.3cm}
\end{table}

{\color{black}
Table~\ref{tab:performance_comparison} collects the main operating point.
The proposed MLR-TG attains the highest raw sum-SE among the implementable codebook-based schemes, at $59.53$~bps/Hz, and the second highest net sum-SE, at $44.89$~bps/Hz, which is $0.96\%$ below GS and $5.48\%$ above AG3.
The reliability metrics separate the schemes far more clearly than the sum rate does.
MLR-TG reduces the system OP by $26.7\%$ with respect to GS, by $26.7\%$ with respect to TG3-TR, by $6.2\%$ with respect to AG3, and by $48.8\%$ with respect to KF-BT, and it raises the $5$th-percentile UE rate by $53.9\%$, $54.8\%$, $14.5\%$, and $167.6\%$ over the same four schemes.
MLR-TG therefore delivers the reliability of always-on deep refinement at the airtime of the greedy baseline, which is the trade-off the policy was calibrated for.
Against the two bounds, MLR-TG is $3.4\%$ below Oracle-L in net sum-SE with an OP within $1\%$ of it, so predicting the level from two scalars instead of evaluating all four costs little, while the remaining gap to Oracle-C, $4.2\%$ in net sum-SE and a factor of $1.8$ in OP, is caused by the candidate restriction and the decision latency rather than by the level rule.
TG3-TR is indistinguishable from GS here, because its single trigger fires in only $4.5\%$ of frames, which is the limitation of fixed-depth triggering that motivated the multi-level design.
}

{\color{black}
Table~\ref{tab:assumptions} compares the computational complexity of the considered schemes with the parameters of Table~\ref{tab:simulation_parameters}.
The per-frame cost shared by every mode amounts to $1.50\times10^{4}$ FLOPs at the UAV, on top of the $1.05\times10^{6}$ FLOPs that the $K$ UEs jointly spend correlating the codebook, which is a receiver-side cost and is identical for all codebook-based schemes.
The coarse stage costs $448$ FLOPs per candidate and the exact predicted-RZF stage costs $1.60\times10^{4}$ FLOPs per promoted subset, so the refinement term reaches $9.71\times10^{5}$ FLOPs at level $3$ against $2.43\times10^{5}$ at level $1$.
With the measured level frequencies, MLR-TG averages $2.31\times10^{5}$ FLOPs per frame, which is $76.6\%$ below the $9.87\times10^{5}$ of AG3 and $15$ times the $1.50\times10^{4}$ of pure GS, and it evaluates $226.8$ candidate subsets per frame on average against $1024$ for AG3.
TG3-TR triggers in $4.5\%$ of frames and therefore averages $5.9\times10^{4}$ FLOPs, which is the arithmetic reason its performance is indistinguishable from that of GS in Table~\ref{tab:performance_comparison}.
OMP-RZF is the most demanding scheme at $8.93\times10^{6}$ FLOPs, dominated by the $N_{\mathrm{RF}}$ correlations of the residual against the whole dictionary.
For reference, the wall-clock time measured for the same four level routines on the simulation host is $18.5$, $789$, $1057$, and $2687~\mu$s, i.e. a factor of $3$ to $5$ above the modeled latency, which is the interpretation overhead of the scripting environment. 
\blue{The modeled latency is used throughout because it represents the implementation-level latency predicted by the stated processor-throughput model, while the scripting-environment wall-clock times are reported only as a reproducibility reference.}
Scaling the modeled latency by $\kappa_{\mathrm{proc}}\in[0.25,8]$, i.e. sweeping the assumed processor from $8$ to $0.25$~GFLOP/s, changes the ranking in the expected direction: AG3 loses $58.3\%$ of its net sum-SE across that range, MLR-TG loses $18.3\%$, and GS loses $7.0\%$.
MLR-TG is the best of the three at $\kappa_{\mathrm{proc}}=0.25$, at $45.89$~bps/Hz against $45.64$ for GS, so the value of adapting the depth grows as the platform becomes more capable and the value of not fixing it at the maximum grows as the platform becomes weaker.
}

\vspace{-0.15cm}
\section{Conclusion}
\label{sec:conclusion}

This paper develops a mobility- and feedback-aware, conflict-triggered HBF framework for MU mmWave UAV systems. 
Building upon the GS baseline and the single-level TG3-TR strategy, the proposed MLR-TG policy selects the refinement depth from both the conflict severity of the reported beam indicators and the expected aging severity, so that deeper candidate refinement is invoked only when its gain justifies its processing cost and delay.
% Starting from the GS baseline and the single-level TG3-TR strategy, the proposed MLR-TG policy generalizes triggered refinement by adaptively selecting the refinement depth according to both the conflict severity of the reported beam indicators and the expected aging severity induced by UAV mobility and feedback delay. In this way, the proposed design preserves low-complexity beam management in ordinary channel realizations while enabling stronger candidate refinement only when the expected performance gain justifies the additional processing effort and delay. 
\blue{Making the framework consistent with the information the transmitter can hold changed what the policy is worth. Under a causal timeline in which every scheme is charged the air time it consumes, the value of adapting the refinement depth appears in the tail of the service distribution rather than in the sum rate: the proposed policy tracks the greedy baseline to within $1\%$ of the net sum-SE while cutting the system outage probability by more than a quarter and lifting the $5$th-percentile UE rate by roughly one half, and it does so while evaluating $77.9\%$ fewer candidate subsets than always-on global top-$3$ refinement.} 
%Predicting the beam direction with a Kalman tracker at the same reporting cost proved less effective than resolving the multiuser conflict, because platform attitude corrections dominate the angular dynamics of a rotary-wing UAV, and the deepest refinement level was selected in fewer than $0.1\%$ of frames, which quantifies how rarely an exhaustive candidate search repays its latency.
\blue{Predicting the beam direction with a Kalman tracker at the same reporting cost proved less effective than resolving the MU conflict in the considered setting, where platform attitude corrections constitute a major component of the angular dynamics of the rotary-wing UAV, and the deepest refinement level was selected in fewer than $0.1\%$ of frames, which quantifies how rarely an exhaustive candidate search repays its latency.}
The UE-scaling and complexity results further confirm a favorable performance-complexity trade-off across heterogeneous operating conditions.
% The ablation, user-scaling, and fairness results further confirmed that the proposed multi-level design provides a favorable performance--complexity trade-off across heterogeneous operating conditions. Several directions deserve further investigation. The proposed framework can be extended to wideband and multi-carrier mmWave UAV systems, more realistic imperfect-feedback settings, and online or learning-based threshold adaptation. It is also of interest to consider multi-antenna users, UAV trajectory-aware beam management, and broader non-terrestrial communication scenarios.
For future research, the extension of the proposed framework to broadband and multi-carrier mmWave UAV systems, online or learning-based threshold adaptation, multi-antenna UEs, and trajectory-aware beam management can be considered.

\balance
\bibliographystyle{IEEEtran}
\bibliography{reference}

\end{document}